\documentclass[preprint2]{aastex62}

\usepackage{amsmath}
\usepackage{color}

\shorttitle{\mosel\ survey: Kinematics of galaxies at $z>3$}
\shortauthors{Gupta et al.}

\newcommand{\mosel}{{\tt MOSEL}}

\newcommand{\zfire}{{\tt ZFIRE}}
\newcommand{\galfit}{{\tt GALFIT}}
\newcommand{\zfourge}{{\tt ZFOURGE}}
\newcommand{\prospector}{{\tt PROSPECTOR}}

\newcommand{\kmos}{{\tt KMOS$^{\rm 3D}$ }}

\newcommand{\oiii}{[\hbox{{\rm O}\kern 0.1em{\sc iii}}]}
\newcommand{\hb}{\hbox{{\rm H}\kern 0.1em{\sc $\beta$}}}
\newcommand{\msun}{M$_{\odot}$}
\newcommand{\logmstar}{$\log(M_*/{\rm M}_{\odot})$}
\newcommand{\sigmaint}{$\log(\sigma_{\rm int})$}

\newcommand{\exsitu}{{\it ex situ}}

\begin{document}

\title{
	
		\mosel\ Survey:  Tracking the Growth of Massive Galaxies at $2<z<4$ using Kinematics and the IllustrisTNG Simulation 
}


\author[0000-0002-8984-3666]{Anshu Gupta}
\affiliation{School of Physics, University of New South Wales, Sydney, NSW 2052, Australia}
\affiliation{ARC Centre of Excellence for All Sky Astrophysics in 3 Dimensions (ASTRO 3D), Australia}

\author[0000-0001-9208-2143]{Kim-Vy Tran}
\affiliation{School of Physics, University of New South Wales, Sydney, NSW 2052, Australia}
\affiliation{ARC Centre of Excellence for All Sky Astrophysics in 3 Dimensions (ASTRO 3D), Australia}
\affiliation{George P.\ and Cynthia Woods Mitchell Institute for Fundamental Physics and Astronomy, Texas A\&M University, College  Station, TX, 77843-4242}

\author[0000-0003-1420-6037]{Jonathan Cohn}
\affil{George P.\ and Cynthia Woods Mitchell Institute for Fundamental Physics and Astronomy, Texas A\&M University, College  Station, TX, 77843-4242}

\author[0000-0002-2250-8687]{Leo Y. Alcorn}
\affiliation{George P.\ and Cynthia Woods Mitchell Institute for Fundamental Physics and Astronomy, Texas A\&M University, College  Station, TX, 77843-4242}
\affiliation{Department of Physics and Astronomy, York University, 4700 Keele St., Toronto, Ontario, Canada, MJ3 1P3}

\author[0000-0002-9211-3277]{Tiantian Yuan}
\affiliation{Swinburne University of Technology, Hawthorn, VIC 3122, Australia}
\affiliation{ARC Centre of Excellence for All Sky Astrophysics in 3 Dimensions (ASTRO 3D), Australia}

\author[0000-0002-9495-0079]{Vicente Rodriguez-Gomez}
\affiliation{Instituto de Radioastronom\'ia y Astrof\'isica, Universidad Nacional Aut\'onoma de M\'exico, A.P. 72-3, 58089 Morelia, Mexico}

\author[0000-0001-9414-6382]{Anishya Harshan}
\affiliation{School of Physics, University of New South Wales, Sydney, NSW 2052, Australia}

\author[0000-0001-6003-0541]{Ben Forrest}
\affiliation{Department of Physics \& Astronomy, University of California, Riverside, 900 University Avenue, Riverside, CA 92521, USA}

\author[0000-0001-8152-3943]{Lisa J. Kewley}
\affiliation{Research School of Astronomy and Astrophysics, The Australian National University, Cotter Road, Weston Creek,
	ACT 2611, Australia}
\affiliation{ARC Centre of Excellence for All Sky Astrophysics in 3 Dimensions (ASTRO 3D), Australia}

\author[0000-0002-3254-9044]{Karl Glazebrook}
\affiliation{Swinburne University of Technology, Hawthorn, VIC 3122, Australia}

\author[0000-0001-5937-4590]{Caroline M. Straatman}
\affiliation{Sterrenkundig Observatorium, Universiteit Gent, Krijgslaan 281 S9, 9000 Gent, Belgium}

\author[0000-0003-1362-9302]{Glenn G. Kacprzak}
\affiliation{Swinburne University of Technology, Hawthorn, VIC 3122, Australia}
\affiliation{ARC Centre of Excellence for All Sky Astrophysics in 3 Dimensions (ASTRO 3D), Australia}

\author[0000-0003-2804-0648]{Themiya Nanayakkara}
\affiliation{Leiden Observatory, Leiden University, P.O. Box 9513, NL 2300 RA Leiden, The Netherlands}

\author[0000-0002-2057-5376]{Ivo Labb\'e}
\affiliation{Swinburne University of Technology, Hawthorn, VIC 3122, Australia}

\author[0000-0001-7503-8482]{Casey Papovich}
\affiliation{Department of Physics and Astronomy, Texas A\&M University, College Station, TX, 77843-4242 USA}
\affiliation{George P.\ and Cynthia Woods Mitchell Institute for Fundamental Physics and Astronomy, Texas A\&M University, College  Station, TX, 77843-4242}

\author[0000-0002-4653-8637]{Michael Cowley}
\affiliation{Centre for Astrophysics, University of Southern Queensland, West Street, Toowoomba, QLD 4350, Australia}
\affiliation{School of Chemistry, Physics and Mechanical Engineering, Queensland University of Technology, Brisbane, QLD 4001, Australia}

\begin{abstract}
We use K-band spectroscopic data from the Multi-Object Spectroscopic Emission Line (\mosel) survey to analyze the kinematic properties of galaxies at $z>3$. Our sample consists of 34 galaxies at $3.0<z_{\rm spec}<3.8$ between $9.0<$\,\logmstar\,$<11.0$.  We find that  galaxies with \logmstar\,$>10.2$ at $z> 3$ have $ 56\,\pm\,21$\,km/s lower integrated velocity dispersion compared to galaxies at $z\simeq 2$ of similar stellar mass. Massive galaxies at $z>3$ have either a flat or declining star formation history (SFH), whereas similar stellar mass galaxies at  $z\sim2.0$ exhibit a slight peak in the past 500 Myrs. Comparing with the IllustrisTNG cosmological simulation, we find that (i) the dynamical mass of massive galaxies in simulations (\logmstar$>10.0$) increases by $\sim0.1$\,dex at a fixed stellar mass between $z=2.0-3.0$, and (ii) dynamical mass growth is coupled with a rapid rise in the \exsitu\ stellar mass fraction (stars accreted from other galaxies) for massive galaxies at $z<3.5$. We speculate that the rising contribution of \exsitu\ stellar mass to the total stellar mass growth of massive galaxies is driving the higher  integrated velocity dispersion and rising SFHs of massive galaxies at $z\sim 2.0$ compared to galaxies of similar stellar masses at $z>3$. 

\end{abstract}


\section{Introduction}

The kinematic properties of star-forming galaxies (SFGs) are intimately related to their mass assembly histories, including both the baryonic (gas and stars) and dark matter assembly. Various spectroscopic surveys have extended our understanding of kinematic evolution of galaxies beyond the local Universe \citep{Epinat2012, Sobral2013, Wisnioski2015, Alcorn2016, Stott2016, Straatman2017, Girard2018}.
The kinematic observation of galaxies at $z>1$  reveals an increasing baryonic fraction of galaxies  with redshift, suggesting an ongoing assembly of dark matter \citep{Gnerucci2011, Lang2016, Genzel2017, Straatman2017, Price2019}.

The mass-assembly history of massive SFGs is different from the assembly history of low mass galaxies.  Massive galaxies (\logmstar\,$\approx\,11$ at $z=0$) acquire almost 40\% of their mass via \exsitu\ processes such as mergers below $z<2$, whereas low mass galaxies mostly grow by {\it in situ} star formation and gas accretion \citep{Nipoti2009, Lee2013, Rodriguez-gomez2016}.  \cite{Rodriguez-gomez2016} using Illustris simulations find that for the most massive galaxies (\logmstar\,$\approx\,12$) at $z=0$, the stellar mass assembly history transitions from {\it in situ} to \exsitu\ growth at around $z\sim 1.0$, whereas the stellar mass assembly of low mass galaxies (\logmstar\,$\approx\,10$ at $z=0$) is dominated by the {\it in situ} star formation at all epochs. 

Observational signatures of transition in the mass assembly histories of massive galaxies are limited. SFGs with compact, dense cores at $z\sim2.0$ are speculated as progenitors of present-day early-type galaxies that transform into elliptical galaxies by addition of \exsitu\ stellar mass from dry mergers \citep{Barro2013, Barro2014, Nelson2014, Wellons2016}. \cite{Nelson2014} and \cite{Barro2014} find compact  SFGs at $z\sim 2.0$  with \logmstar\,$ \sim 10.8-11.0$ have integrated velocity dispersions nearly equal to the stellar velocity dispersions of massive quiescent galaxies at $z\sim 2.0$. 

In this paper, we show that kinematic properties of galaxies at $z>3$ is consistent with the transitory phase ({\it in situ} to \exsitu\ growth) in the assembly history of massive galaxies between $z=2-3$.  Current investigations into the kinematics of galaxies at $z>3.0$ are limited by the small number of galaxies, especially with \logmstar$\,>10.0$ \citep{Law2009, Gnerucci2011, Livermore2015, Turner2017, Girard2018, Price2019}.  We use the K-band spectroscopic data from the Multi-Object Spectroscopic Emission Lines (\mosel) survey (Tran et al., submitted).  We derive SFHs of our \mosel\ targets using  the spectral energy distribution fitting code \prospector\ \citep{Leja2017} and  compare our results with the dynamical mass estimates from the IllustrisTNG simulations \citep{Pillepich2017a, Nelson2017}.

\begin{figure*}
	\centering
	\tiny
	\includegraphics[scale=0.37, trim=0.0cm 0.0cm 0.0cm 0.0cm,clip=false]{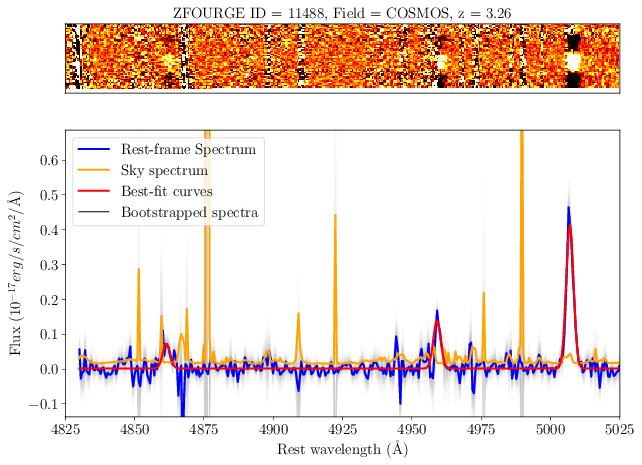}
	\includegraphics[scale=0.37, trim=0.0cm 0.0cm 0.0cm 0.0cm,clip=false]{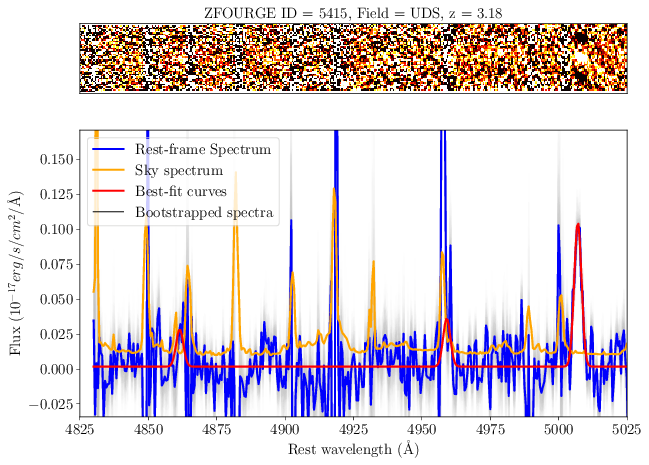}
	\caption{Sample spectra for two \mosel\ galaxies from MOSFIRE observations. The blue curves are the observed 1D spectra in the rest frame, orange is corresponding noise spectra, and the gray shaded region is the bootstrapped iterations. The red curves correspond to the best-fit curves to the \oiii${\lambda\lambda5007,4959}$ and \hb${\lambda4861}$ emission lines. The image in top panel shows the corresponding 2D spectrum for each galaxy. We provide MOSFIRE spectra of all massive \mosel\ galaxies in Apppendix \ref{fig:spectra_out}.}
	\label{fig:spectra}
\end{figure*}

This paper is organized as follows. We discuss our methodology and observations in Section \ref{sec:obs}. In Section \ref{sec:mosel_survey}, we  describe the sample selection, observation and data reduction for the  \mosel\ survey.   Section \ref{sec:kinematics} presents our results from observations. In Section \ref{sec:simulations}, we compare our results with IllustrisTNG simulations. Finally, in Section \ref{sec:discussion} we discuss the main implications of our results and summarise them in Section \ref{sec:summary}.

For this work, we assume a flat $\Lambda$CDM cosmology with $\Omega_{M}$=0.3, $\Omega_{\Lambda}$=0.7, and $h$=0.7. The only exception is the \prospector\ software, where a WMAP9 cosmology \citep{Hinshaw2013} cosmology is used.

\begin{figure}
	\centering
	\tiny
	\includegraphics[scale=0.35, trim=0.5cm 0.5cm 0.0cm 0.0cm,clip=true]{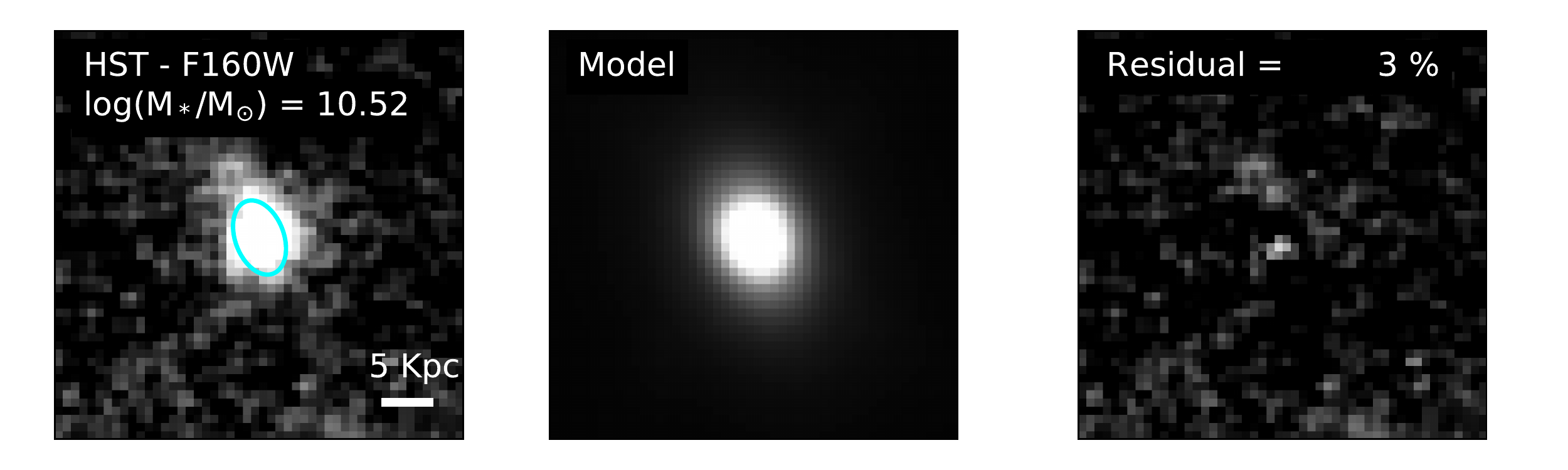}
	\caption{HST-F160W image (left) of a sample galaxy from the \mosel\ survey. The middle and the right panels show the best-fit spatial profiles and residuals, respectively from \galfit.  The cyan ellipse in the left panel indicates the best-fit ellipse along the semimajor axis with a radius  equal to the  twice the effective radius. The legend in the right panel indicates the percentage residuals summed in quadrature within twice the effective radius. For velocity dispersion and dynamical mass analysis, we select galaxies with the total percentage residuals $<20\%$.  We provide \galfit\ models for all massive \mosel\ galaxies in Appendix \ref{fig:galfit_out_massive}. }
	\label{fig:galfit_out_ex}
\end{figure}

\section{Observations}\label{sec:obs}

\subsection{\mosel\ survey}\label{sec:mosel_survey}
Our sample is drawn from the \mosel\ survey, which is a spectroscopic follow-up of the $z\sim 3$ galaxies selected from the FourStar Galaxy Evolution survey \citep[\zfourge;][]{Straatman2016}. The \zfourge\ survey uses the medium $J-$band filters $J_1$, $J_2$, and $J_3$, and medium $H-$band filters $H_s$ and $H_l$, and deep $K_s$ filters, to target specific spectral features for galaxies at $2.5<z<4$.  Thus, \zfourge\ survey reaches a photometric redshift accuracy of  $\sigma_z = 0.016$  in the redshift range $2.5<z<4.0$ \citep{Straatman2016}.  The \zfire\ survey confirms the precision of the photometric redshift measurement of the \zfourge\ survey to $\sigma_z \sim 2\%$ \citep{Nanayakkara2016}.

We refer to  Tran et al.\,(submitted) for a detailed description of the \mosel\ survey design. In summary, the \mosel\ survey acquires near-infrared spectra of the emission line galaxies between redshift $3.0<z<3.8$ to understand their contribution to the star formation history of the universe.  The \mosel\ survey uses the emission line strength defined as  [OIII]+H$\beta$ equivalent widths (EW) from the composite spectral energy distributions (SEDs) fitting by \cite{Forrest2018} to identify emission line galaxies. Based on the strength of the [OIII]+H$\beta$ EW, the \mosel\ survey classifies emission line galaxies at $2.5<z<4$ as follows: extreme emission line galaxies with [OIII]+H$\beta$ EW $>800$\AA, strong emission line galaxies with [OIII]+H$\beta$ EW $230-800$\AA, and star-forming galaxies with [OIII]+H$\beta$ EW $0-230$\AA.  

Throughout this paper, we use stellar masses derived in the \zfourge\ survey using the SED fitting code FAST \citep{Kriek2009} for both the \mosel\ and \zfire\ galaxy samples.

\begin{figure}
	\centering
	\tiny
	\includegraphics[scale=0.35, trim=0.5cm 0.5cm 0.5cm 0.5cm,clip=true]{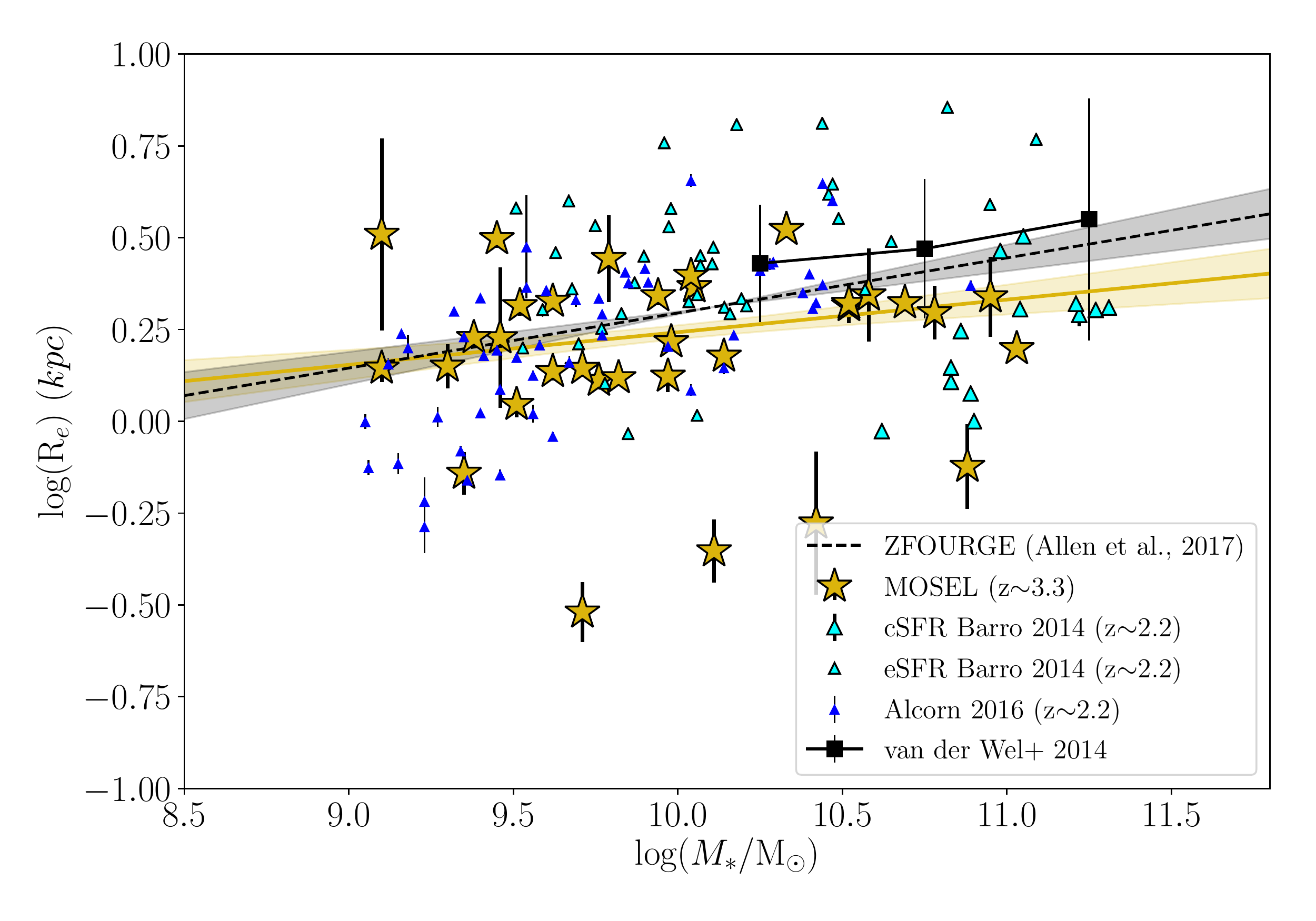}
	\caption{The stellar mass versus effective F160W radii ($R_e$) relation for \mosel\ galaxies (gold stars), where $R_e$ was derived using \galfit. The gold shaded region is the best-fit linear relation between the $\log(R_e)$ and stellar mass. The black dashed line is stellar mass-size relation for galaxies between $z=3-4$ from \zfourge\ survey by \cite{Allen2016}.  The black squares show the median and 16$^{th}$ and 84$^{th}$ percentile in the size of late type galaxies at z=2.75 from \cite{VanderWel2014}.  We compare our measurements with various star-forming galaxy samples at $z\sim2$, \citep{Alcorn2016, Barro2014}. }
	\label{fig:mass_re}
\end{figure}

\subsection{Keck/MOSFIRE observations}\label{sec:mosfire}
Keck/MOSFIRE \citep{McLean2012} observations were taken on 12 and 13 February 2017 (project code Z245, PI Kewley). A total of 5 masks were observed in COSMOS field and 1 mask in CDFS field in $K$-band filter covering a wavelength of $1.93-2.38\,\micron$. The spectral dispersion is 2.17\,\AA/pixel. The seeing was $\sim 0.7''$. 

A total of 95 galaxies were targeted between $0.9<z<4.8$, with highest priority given to the emission line galaxies with  \oiii+\hb equivalent width  $>230$\,\AA\ (38 galaxies) between $2.5<z<4.0$. Possible active galactic nuclei (AGN) contaminants were removed using the \cite{Cowley2016} catalog that uses X-ray, radio, and infrared imaging to identify AGNs in the \zfourge\ survey. The data was reduced using the MOSFIRE data reduction pipeline\footnote{http://keck-datareductionpipelines.github.io/MosfireDRP} and flux calibration was performed using the \zfire\ data reduction pipeline \citep{Tran2015, Nanayakkara2016}. 

We spectroscopically confirm 48 galaxies between $2.9<z<3.8$ of which 11 are extreme emission line galaxies, 13 are strong emission line galaxies, and 24 are star-forming galaxies (Tran et al. submitted). We also add data for $z>3.0$ galaxies observed in the \zfire\ survey. The median redshift of our sample is $z_{\rm spec}= 3.4$.  We reach a final sample of 34 galaxies after selecting galaxies with signal-to-noise (S/N) greater than three (see Section \ref{sec:em_flux}) and \galfit\ residuals $<20\%$ (see Section \ref{sec:hst_imaging}). 

\subsection{Emission line flux and kinematic measurements}\label{sec:em_flux}

We use the flux calibrated and telluric corrected 2D slit spectra from the \mosel\ survey and and $z>3$ galaxies from the \zfire\ survey \citep{Tran2015, Nanayakkara2016} to extract emission line fluxes. We collapse the 2D slit spectra along the wavelength axis to generate the spatial profile and fit a Gaussian. To generate the 1D spectra, we sum the 2D slit spectra within two times the full-width half maximum (FWHM) from the centroid of the spatial profile. To generate an error 1D spectrum, we sum the noise 2D slit spectrum in quadrature within the same aperture as the 1D flux spectrum.  

We initally manually identified the line centroid to provide an initial galaxy redshift. This was possible given the high S/N of the emission lines. We then deredshifted the spectra and computed the final glaaxy redshifts along with the emission line fluxes after performing a Gaussian fit to emission lines. We simultaneously fit the \oiii\,$\lambda 5007$, \oiii\,$\lambda 4959$ and \hb\,$\lambda 4861$ emission lines with three Gaussians and five free parameters: redshift, flux-\oiii, flux-\hb, width, and continuum level. We fix  the \oiii\,$\lambda 4959$ flux to be \oiii\,$\lambda 5007$/3. For galaxies where \hb\ S/N is $<3$, we refit the 1D-spectrum including only \oiii\,$\lambda 5007$ and \oiii\,$\lambda 4959$ emission lines with two Gaussians and four free parameters: redshift, flux-\oiii, width, and continuum level.

The instrumental broadening is measured from the width  of the skylines in K-band in the error spectrum near 5007\,\AA\ in wavelength units and is 0.55 \AA\, (32 $\rm{km\, s}^{-1}$). While fitting emission lines, we subtract the instrumental broadening in quadrature from the Gaussian line width.  For galaxies where only a single emission line was detected, we assume that emission line identification is  correct if the difference between \zfourge\ photometric redshifts and spectroscopic redshift is  less than $2\%$ \citep{ Tran2015, Nanayakkara2016}.

Even in the best seeing conditions ($0.5''$) galaxies at $z\sim 3$ cannot be resolved with MOSFIRE. We resort to using the integrated velocity dispersion ($\sigma_{int}$) measured using the integrated line width to estimate the kinematic properties of galaxies. 
We determine the integrated velocity dispersion using the best-fit line width to the highest S/N line \oiii\,$\lambda 5007$. 

Figure \ref{fig:spectra} shows two randomly selected sample spectra.  For each galaxy, we create 1000 realization of the flux spectrum by perturbing the flux spectrum according to the noise spectrum (shown as a gray shaded region in Figure \ref{fig:spectra}). For each realization, we perform the previously described fitting routine to remeasure the emission line fluxes and the instrumental dispersion corrected integrated velocity dispersion ($\sigma_{int}$).  The standard deviation from the bootstrapped versions represents the noise in the line flux and line width measurements. All our results have been quoted with at least \oiii\,$\lambda 5007$ detections at a S/N greater than 3.  

\begin{figure}
	\centering
	\tiny
	\includegraphics[scale=0.45, trim=0.0cm 0.0cm 0.0cm 0.0cm,clip=true]{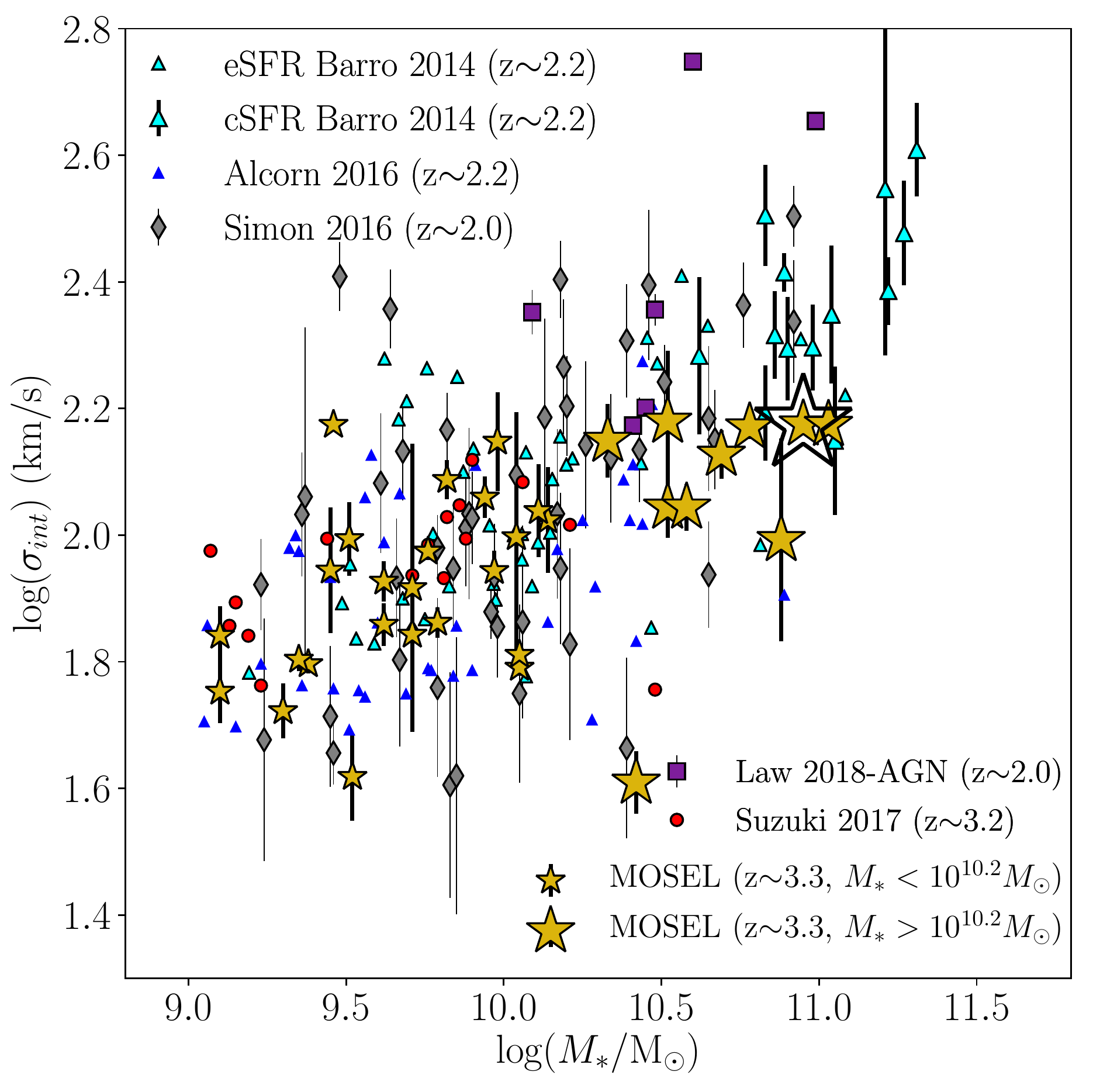}
	\caption{{\bf Left :} Distributions of the integrated velocity dispersion (\sigmaint) versus stellar mass  for the \mosel\ galaxies (gold stars). The small and large gold stars correspond to  the  two sub-groups identified using the k-Means clustering algorithm separated at a stellar mass of  $M_* = 10^{10.2}$\msun.   The big open star corresponds to a galaxy with broad emission features indicative of an AGN.  We compare the \mosel\ sample with SFGs at $z\sim 2$ from \cite{Barro2014}, \cite{Alcorn2016}, and \cite{Simons2016}. Additionally, we compare with SFGs at $z\sim 3$ from \cite{Suzuki2017} and AGNs at $z\sim 2$ from \cite{Law2018}.  }
	\label{fig:st_mass_vel}
\end{figure}

\begin{figure*}
	\centering
	\tiny
	\includegraphics[scale=0.33, trim=0.0cm 0.0cm 0.0cm 0.0cm,clip=true]{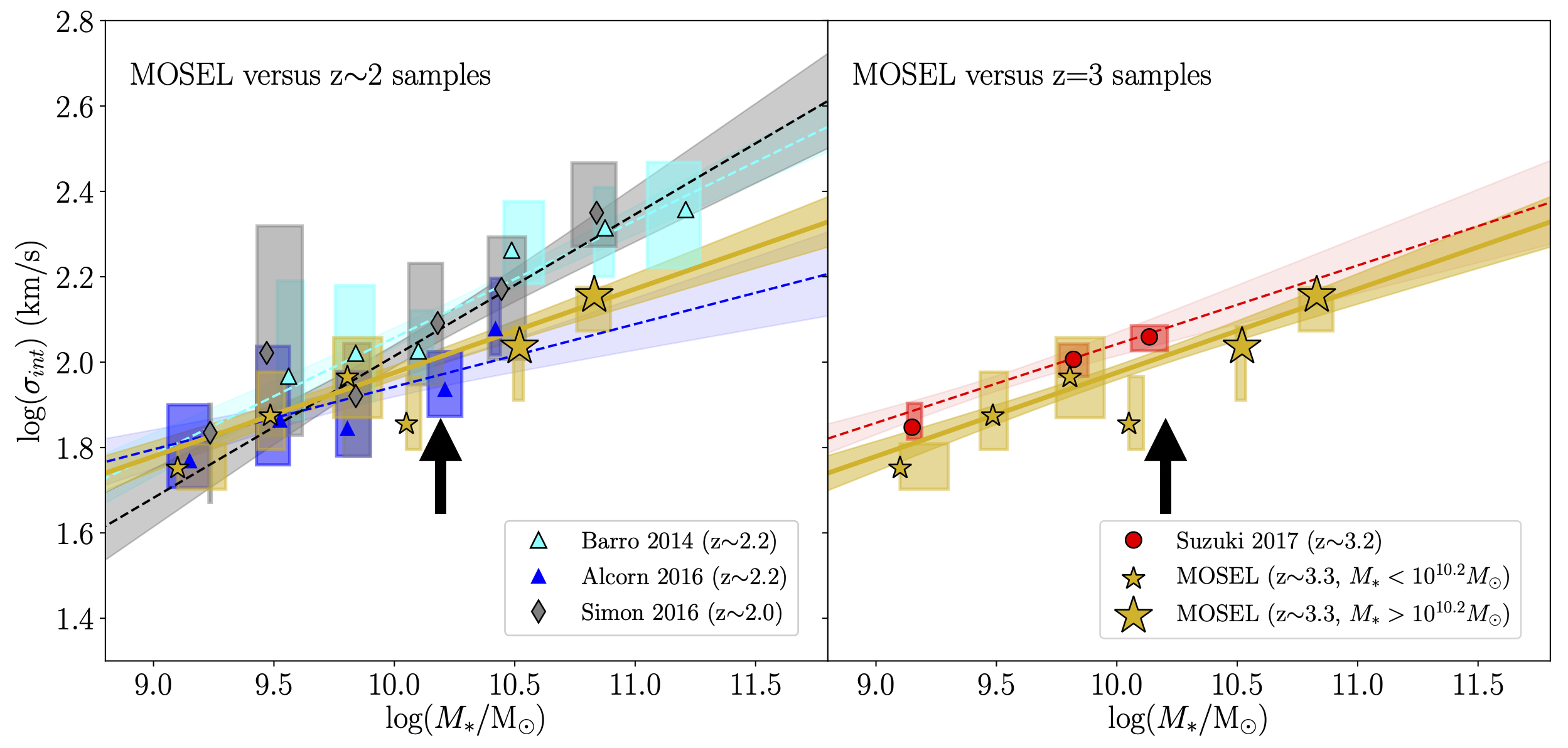}
	\caption{  The Integrated velocity dispersion (\sigmaint) for the \mosel\ galaxies (gold stars) binned into stellar mass in comparison with the sample at $z\sim2$ (left) and $z>3 $ (right). The color scheme is the same as in Figure \ref{fig:st_mass_vel}. The symbols and the shaded rectangles correspond to the median and the 25$^{\rm th}$ to 75$^{\rm th}$ percentile in \sigmaint\ and stellar mass after bootstrapping  respectively.   The dashed lines represent the best-fit linear relation between $M_*$ and \sigmaint\ for the respective sample.  Massive galaxies with \logmstar$>10.2$ (pointed by the black arrow) from our \mosel\ survey have lower \sigmaint\ compared to galaxies  of similar stellar masses  at $z\sim2$. }
	\label{fig:st_mass_vel_binned}
\end{figure*}

\subsection{HST imaging}\label{sec:hst_imaging}
To measure effective radii of galaxies, we use Cosmic Assembly Near-Infrared Deep Extragalactic Survey \citep[CANDELS;][]{Koekemoer2011,Grogin2011} imaging. We use the composite PSF images for the HST filters from the 3D-HST survey \citep{Brammer2012}. We measure the effective radius (${\rm R_e}$) using the \galfit\ software \citep{Peng2010a} and HST-F160W images. 

We fit a single-S\'ersic profile to galaxies, with initial parameters for the disk size, axis ratio, and position angles taken from the \cite{VanderWel2014} and visual inspection. Only $\sim$20\% of our \mosel\ targets had effective radius measurements in the original  \cite{VanderWel2014} catalog. We ran \galfit\ on our galaxies in an automated fashion and visually inspected the residual images to determine the goodness of fit.  We use the following constraints during the automated S\'ersic profile fitting: centroid $\Delta x = \pm 3\,{\rm pixels}, \Delta y = \pm 3\,{\rm pixels}$, S\'ersic index = $0.7-4$, $\Delta {\rm R_e} = \pm 3$\,pixels and $\Delta$position angle $=\pm 10^{\circ}$. Galaxies with poor fits were refitted by varying the  initial parameters till a good fit was obtained or the galaxy was deemed to have too low S/N for a reasonable fit. 

 We use the \galfit\ best-fit parameters to  determine the sizes,  axis ratios and position angles of galactic disks on the sky. We measure the residual fraction for each galaxy  by summing the residual image in quadrature within the galactic disk and dividing by the total flux within the same region. Throughout this paper, we use galaxies that have total residual flux after surface brightness fitting less than 20\%. 

Figure \ref{fig:galfit_out_ex} shows an example of the surface brightness profile fit for a randomly selected galaxy.   Figure \ref{fig:mass_re} shows the derived relation between stellar mass and the semi-major axis  radii  from \galfit\ as effective radii for our \mosel\ targets.  We find that our effective radii measurements are within 1-sigma error compared to effective radii measurements  in the \cite{VanderWel2014} catalog. 
We note that our best-fit stellar mass-size relation for the \mosel\ sample is slightly below the 1-sigma stellar mass-size relation derived by \cite{Allen2016} for $z=3-4$ galaxies from the \zfourge\ survey.  However,  there is a lot of scatter in the mass-size relation at $z>3$ at the massive end due to small number statistics. 

\subsection{Star formation histories from \prospector}\label{sec:prospector}

\prospector\ uses a Flexible Stellar Population Synthesis package \citep[FSPS;][]{Conroy2009, Conroy2010} where the contribution of dust attenuation, nebular emission, and re-radiation was modeled \cite{Byler2017}.  We use the parameters used by \cite{Cohn2018} to the SFHs, i.e., using a \cite{Chabrier2003} IMF with MESA Isochrones \& Stellar Tracks \citep[MIST;][]{Dotter2016, Choi2016, Paxton2011, Paxton2013, Paxton2015}, \cite{Calzetti1994} dust attenuation model with a WMAP9 cosmology \citep{Hinshaw2013}. We fit nine free parameters, stellar mass, stellar and gas-phase metallicity, dust attenuation and five independent nonparametric SFH bins.

\prospector\ fits non-parametric SFHs by fitting the fraction of stellar mass formed in a particular time bin, after fitting for the total stellar mass \citep{Leja2019}. To isolate the emission from young and old stars, we used the following time bins: 0-50 Myr, 50-100 Myr, 100-500 Myr, 0.5-1.0 Gyr, and two evenly spaced time bins from 1 Gyr to the age of the Universe at the redshift of a given galaxy. \prospector\ fits for 6 SFH bins  but the additional constraint on fractional stellar mass to be one results in only five independent SFH bins. We use a uniform prior on SFH  corresponding to a constant star formation rate.

\begin{table*}
	\small
	\begin{center}
		\caption{Best lest-square linear fits to the integrated velocity dispersion and stellar mass distribution for $z>2$ observations}
		\label{tb:best_fit_vel_dispersion}
		\begin{tabular}{ p{7.2 cm}  p{2.0cm}  p{2.0cm} p{1.0cm}  p{2.5cm}  }
			\hline
			\hline
			Sample & \sigmaint\ [km/s]$^a$ & Slope$^b$ & $N^c$ & $\log(M_{dyn}/{\rm M}_{\odot})^a$  \\
			\hline
			
			\cite{Barro2014}, $z\sim2.0$ & $1.74\pm0.04$ & $0.31\pm0.03$ & 53 & $9.46\pm0.08$ \\
			\cite{Barro2014}, \logmstar$<10.2$ & $1.91\pm0.10$ & $0.12\pm0.1$ & 30 & $9.52\pm0.06$ \\
			\cite{Barro2014} \logmstar$> 10.2$ & $1.70\pm0.20$ & $0.32\pm0.1$ & 23 & $9.21\pm0.07$ \\
			\cite{Alcorn2016}, $z\sim2.0$ & $1.80\pm0.05$ &$0.15\pm0.05$ & 41 & $9.25\pm0.05$\\
			\cite{Alcorn2016}, \logmstar$<10.2$ & $1.80\pm0.06$ &$0.14\pm0.08$ & 30 & $9.41\pm0.05$\\
			\cite{Alcorn2016}, \logmstar$>10.2$ & $1.94\pm0.48$ &$0.05\pm0.32$ & 11 & $9.14\pm0.13$\\
			\cite{Alcorn2016}, size evolution to $z=3$  & $1.65\pm0.05$ &$0.26\pm0.05$ & 41 & $9.14\pm0.05$\\
			\cite{Simons2016}, $z\sim 2.0$ & $1.68\pm0.07$ & $0.33\pm0.06$ & 48 & ......... \\
			\cite{Suzuki2017}, $z\sim3.0$ & $1.86\pm 0.03$ & $0.17\pm0.04$ & 17 &  ......... \\
			\mosel\ (this work), $z\sim3.0$ & $1.78\pm 0.04$ & $0.19\pm 0.03$ & 34 & .......... \\
			\mosel\ (this work, \logmstar$<10.2$) & $1.74\pm 0.06$ & $0.23\pm 0.07$ & 24 & $9.39\pm 0.08$\\
			\mosel\ (this work, \logmstar$>10.2$) & $2.06\pm 0.21$ & $0.03\pm 0.11$ & 10 & $8.75\pm 0.19$ \\			
			\hline 
			\hline
		\end{tabular}
	\end{center}
	
	\begin{flushleft}
		{\bf Notes:}\\
		$^{\rm a}$ at \logmstar$=9$ from the best linear fit. \\
		$^{\rm b}$ slope of best-fit relation of the form \sigmaint$ =A+ B*$\logmstar. \\
		$^{\rm c}$ Number of objects used for linear fit.\\
		 
	\end{flushleft}
	
\end{table*}%

\section{Kinematics and SFHs of  galaxies at $z>3$}\label{sec:kinematics}

\subsection{Kinematics of \mosel\ galaxies}\label{sec:velocity_dispersion}

 The integrated velocity dispersion represents a combination of the rotation and the intrinsic velocity dispersion of galaxies \citep{Glazebrook2013, Barro2014}. 
 We use the integrated velocity dispersion from the \oiii\ emission line to analyse evolution of the gravitational potential and the intersic velocity dispersion of  $z > 3$ galaxies from our \mosel\ survey.  Resolving kinematics for $z\sim 3$ galaxies with MOSFIRE  is not possible because of the small disk size of $z> 3$ galaxies and seeing-limited conditions. 

Figure \ref{fig:st_mass_vel} shows  \sigmaint\ as a function of the stellar mass for \mosel\ galaxies.  We select galaxies with S/N$>3$ on \oiii\,$\lambda 5007$ (Figure \ref{fig:spectra}) and small \galfit\ ($<20\%$) residuals (Figure \ref{fig:galfit_out_ex}).  The limited spectral and spatial resolution of MOSFIRE at $z\sim3.0$ prevents full kinematic decomposition of galaxies similar to \cite{Straatman2017}, restricting us to \sigmaint\ measurements.

\subsubsection{Selecting the stellar mass cut-off}

We use k-Means,  a python based unsupervised learning clustering algorithm by the scikit-learn \citep{Pedregosa2012} library, to identify two sub-groups of \mosel\ galaxies on the \sigmaint\ versus stellar mass plane, separated at a stellar mass of \logmstar\,$= 10.2$. The two identified sub-groups have the locations, g1: \logmstar\,$=9.7$, \sigmaint\,$=1.91$ and g2: \logmstar\,$=10.7$, \sigmaint\,$=2.05$. The stellar mass cutoff \logmstar\,$=10.2$  identified by the k-Means algorithm is similar to the turnover stellar mass in the stellar mass function at $z > 2.0$ \citep{Davidzon2017}.	  In the rest of the paper, we use the \logmstar\,$=10.2$ cut-off to separate galaxies into low and high stellar mass bins.

\subsubsection{Comparison with $z> 2$ samples from literature}

We compare the kinematic properties of \mosel\ galaxies with other slit-based studies at $z\sim2.0$ \citep[Figure \ref{fig:st_mass_vel};][]{Barro2014, Alcorn2016, Simons2016, Suzuki2017}. We select the \cite{Simons2016} and \cite{Barro2014}  samples because they cover the full stellar mass range of our \mosel\ sample. From \cite{Barro2014} we combine  samples of both compact and extended SFGs because they exhibit a similar relation between the stellar mass and \sigmaint.

 To derive \sigmaint\ for the \cite{Simons2016} sample, we add the  inclination-uncorrected rotation velocity and intrinsic velocity dispersion in quadrature. Although our derived  \sigmaint\ from the \cite{Simons2016} sample does not equal the observed \sigmaint, we infer from the rotational model that  the difference would be  smaller than 20km/s, significantly smaller than our measurement errors. We also compare with the \cite{Alcorn2016} sample, which extends up to \logmstar\,$\leq 10.5$. The stellar masses for all comparison samples  are derived using FAST \citep{Kriek2009}. 
 
We use the python package Lmfit \citep{Newville2016} to fit a linear relation of the form \sigmaint\,$ = A + B \times \log(M*/M_{\odot})$ to the stellar mass and \sigmaint\ distribution for the various samples after running an iterative  2.5$\sigma$ outlier rejection. Table \ref{tb:best_fit_vel_dispersion} shows the best-fit parameters of the linear relation for various comparison samples.  The quoted uncertainties in Table \ref{tb:best_fit_vel_dispersion} are derived from the covariance matrix, which are consistent within $1\sigma$ to the uncertainties derived via bootstrapping.  We find that massive galaxies (\logmstar\,$ > 10.2$) at $z\sim3$ have lower \sigmaint\ compared to galaxies of similar stellar masses at $z\sim2$.  

 We bin the \sigmaint\ measurements for the \mosel\ and various comparison samples in stellar mass  (Figure \ref{fig:st_mass_vel_binned}). We require the stellar mass bins to have at least 2 galaxies in the respective sample. For each sample, we create 100 realisations for all galaxies in a particular stellar mass bin by perturbing the data points according to their uncertainties.  We estimate a median and the area corresponding to the 25$^{\rm th}$ to 75$^{\rm th}$ percentile in the stellar mass and \sigmaint\ for each stellar mass bin.

The \sigmaint\ and stellar mass distribution for the low mass \mosel\ sample (\logmstar\,$< 10.2$) is consistent with other studies at $z\sim2$ (Figure \ref{fig:st_mass_vel_binned}: left panel). The SFGs from \cite{Suzuki2017} have nearly 0.1\,dex higher \sigmaint\ compared to the \mosel\ sample (Figure \ref{fig:st_mass_vel_binned}: right panel). We suspect that the selection of galaxies via narrow-band imaging biases the \cite{Suzuki2017} sample towards the high specific star formation rate (sSFR) galaxies. An intrinsic bias toward high sSFR might result in the selection of galaxies with high intrinsic velocity dispersion \citep{Ubler2019}, and in turn high \sigmaint. 
 
Figure \ref{fig:st_mass_vel_binned} shows that the \mosel\ galaxies with \logmstar\,$>10.2$ have lower \sigmaint\ compared to the same stellar mass galaxies $z\sim2.0$. We cannot compare the \sigmaint\ measurements for the massive \mosel\ galaxies with the \cite{Suzuki2017} sample because they only have  one galaxy with \logmstar$>10.0$. We combine the integrated velocity dispersion measurements for massive galaxies (\logmstar\,$>10.2$) in \cite{Barro2014} and \cite{Simons2016}, and bootstrap to estimate a median \sigmaint\ of massive galaxies at $z\simeq2$. After bootstrapping the \sigmaint\ for the massive \mosel\ galaxies, we estimate that massive \mosel\ galaxies have nearly $56\,\pm\,21$\,km/s lower integrated velocity dispersion compared to the similar stellar mass galaxies at $z\sim2.0$.


\subsubsection{Contamination from AGN and mergers}
 
To identify the role of AGNs, we compare our measurements with a sample of 6 narrow-line AGNs from \cite{Law2018}. The integrated velocity dispersion of the \cite{Law2018} sample shows a large scatter, where only two galaxies have  $\sigma_{\rm int}\,>500\,$km/s. The \cite{Barro2014} sample might also have some contribution from AGNs, because some of their compact star-forming galaxies exhibit broad emission features with X-ray emission. To remove AGNs from the  \mosel\ sample, we use  X-ray, radio, and infrared emission catalogs from \cite{Cowley2016}. One of  our massive \mosel\ galaxies shows a clear sign of broad emission, indicative of either AGN, shocks or outflows.  We do not rule out contamination from narrow-line AGNs in our massive galaxy sample.
 
 Our results cannot be explained by the higher probability of misclassification of mergers as rotating disks at $z\sim3.0$ compared to $z\sim 2.0$ \citep{Hung2015a}. We do find some indications of extended diffuse components for some galaxies (Figure \ref{fig:galfit_out_massive}), indicative of mergers. However, misclassified mergers as rotating disks at $z\sim3.0$ would result in a relatively higher observed \sigmaint\ for galaxies at $z\sim 3.0$, in contrast to our result. 

We speculate that the lower integrated velocity dispersions we find for massive galaxies at $z>3$ as compared to similar mass galaxies at $z\sim2$ indicates either the rotation velocity or intrinsic velocity dispersion decreases for massive galaxies from redshift 2 to 3 (See Section \ref{sec:discussion}).

\begin{figure*}
	\centering
	\tiny
	\includegraphics[scale=0.32, trim=0.0cm 0.0cm 0.0cm 0.0cm,clip=true]{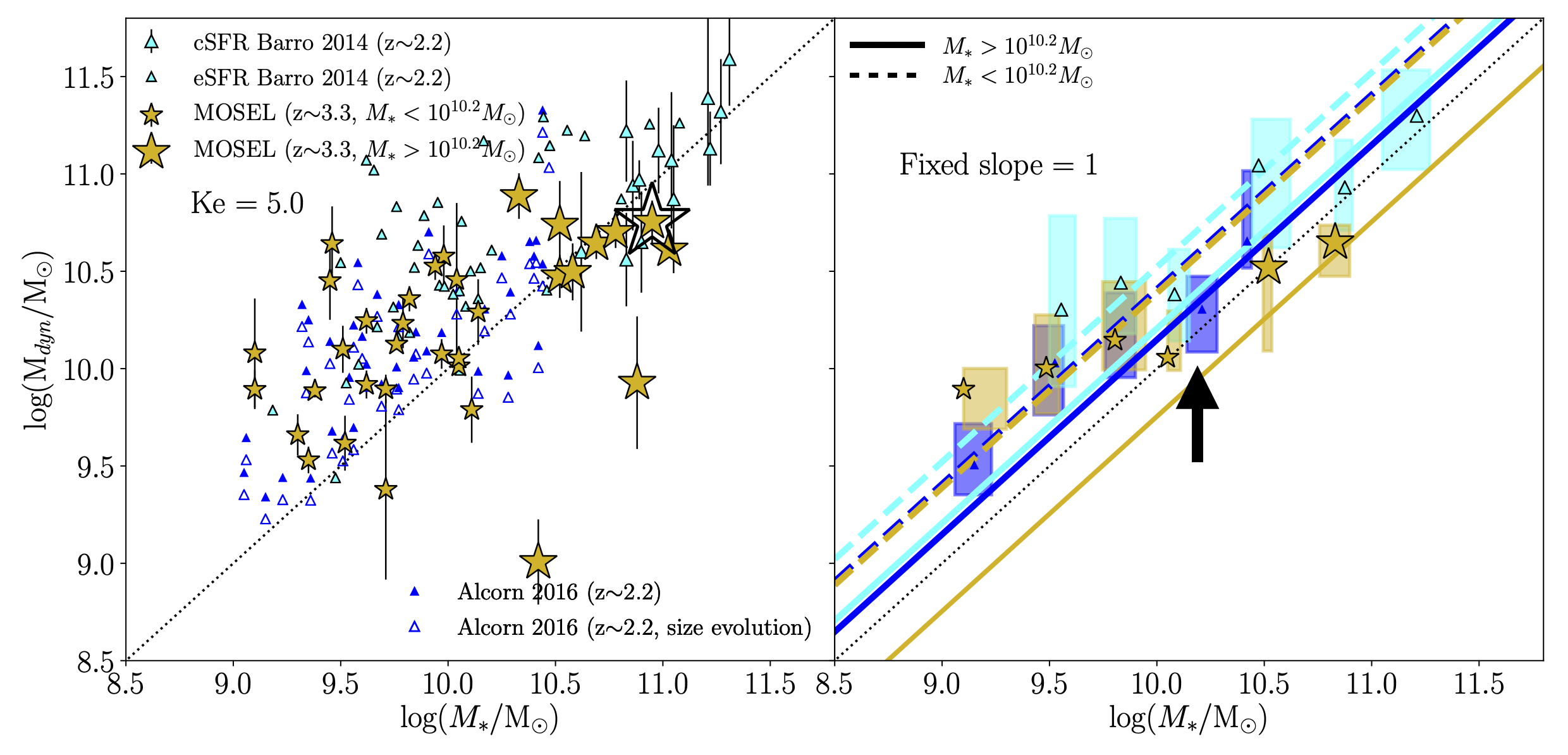}
	\caption{ Comparison between the dynamical mass and stellar mass in the left panel.  The color scheme is the same as in Figure \ref{fig:st_mass_vel}. The open blue triangles in the left panel correspond to the size-evolution corrected dynamical mass estimates for the  \cite{Alcorn2016} sample.  The right panel shows the dynamical masses after stellar mass binning each sample. The symbols and shaded rectangles correspond to the median and 25$^{\rm th}$ to 75$^{\rm th}$ percentile.  The colored dashed and solid lines in the right panel correspond to the best-fit dynamical mass for galaxies with \logmstar$<10.2$ (pointed by the black arrow) and \logmstar$>10.2$ respectively for each sample. We find that massive galaxies at $z\sim3.0$ have a lower dynamical mass compared to  $z\sim2$ galaxies  of similar stellar masses. }
	\label{fig:dynamical_mass}
\end{figure*}

\subsection{Dynamical mass Analysis}\label{sec:dy_mass}

Figure \ref{fig:dynamical_mass} shows the comparison between the dynamical mass and the stellar mass for \mosel\ galaxies. To measure dynamical masses, we use the virial theorem  
\begin{equation}\label{eq:virial_theorm}
    M_{dyn} = Ke\frac{\sigma_{\rm int}^2R_e}{G}
\end{equation}

where $R_e$ is the effective radius and $Ke$ is the virial factor. The effective radius was measured using  \galfit\ on HST-F160W imaging (Section \ref{sec:hst_imaging}). 
The value of $Ke$ depends on the mass profile, the ratio of velocity dispersion to the rotation, and the shape of the overall gravitational potential \citep{Courteau2014}. The virial factor can range between $2-10$ depending on the overall structure of the galaxy \citep{Maseda2013, VandeSande2013}.  

To consistently compare with studies at $z>2$, we choose a virial factor of $Ke = 5$, which is typically used for dispersion dominant disks at high redshifts \citep{Barro2014, Maseda2014, Alcorn2016, De2014}. The choice of the virial factor would change our dynamical mass estimates but does not affect the main conclusion of this paper. We compare the dynamical mass estimates for \mosel\ galaxies with the \cite{Alcorn2016} and \cite{Barro2014} samples because they provide effective radii or dynamical masses for their sample.   We determine the offset between the dynamical mass and stellar mass for various observational studies by performing a linear fit with $2.5 \sigma$ outlier rejection at a fixed slope of 1 (Table \ref{tb:best_fit_vel_dispersion}).  For a reliable estimate of effective radii, we only select \mosel\ galaxies where residuals after surface brightness profile fitting via \galfit\ are less than 20\%.   

We estimate that massive galaxies (\logmstar\,$>10.2$)  in our \mosel\ sample have nearly 0.4\,dex lower dynamical mass compared to galaxies of similar stellar masses at $z\sim 2$. The relation between dynamical mass and stellar mass for the low mass \mosel\ sample (\logmstar\,$<10.2$) is consistent within 1-sigma errors to other studies $z>2.0$. The two massive galaxies with un-physical dynamical masses in our sample are extremely compact  (see Figure \ref{fig:mass_re}), and one of them has integrated velocity dispersion close to the spectral resolution limit of MOSFIRE ($\sim 32$\,km/s).   By analyzing the inclination of massive galaxies on sky from the \galfit, we rule-out a preference towards face-on galaxies in our \mosel\ sample (Figure \ref{fig:galfit_out_massive}).


The stellar mass to effective radii distribution of our \mosel\ sample is similar to the \cite{Barro2014} and  \cite{Alcorn2016} samples (Figure \ref{fig:mass_re}). The effective radii of our galaxies are consistent within 1$\sigma$ errors to the stellar mass versus size relation derived by \cite{Allen2016} using the \zfourge\ data for galaxies at $z=3-4$ (Figure \ref{fig:mass_re}).   We estimate that a simple size evolution of galaxies between $z=2$ to $z=3$ would result in the observation of a 0.1\,dex lower dynamical mass measurement for galaxies at $z=3$ (open blue triangles in Figure \ref{fig:dynamical_mass}: left panel). However, an offset of $0.1\,$dex is insufficient to explain the $\sim 0.5$\,dex lower dynamical mass of massive galaxies at $z\sim 3$ compared to galaxies  of similar stellar masses at $z\sim2$. We speculate a shift in the evolutionary pathway of massive galaxies between $z=2-3$. 

\begin{figure*}
	\centering
	\tiny
	\includegraphics[scale=0.32, trim=0.0cm 0.0cm 0.0cm 0.0cm,clip=true]{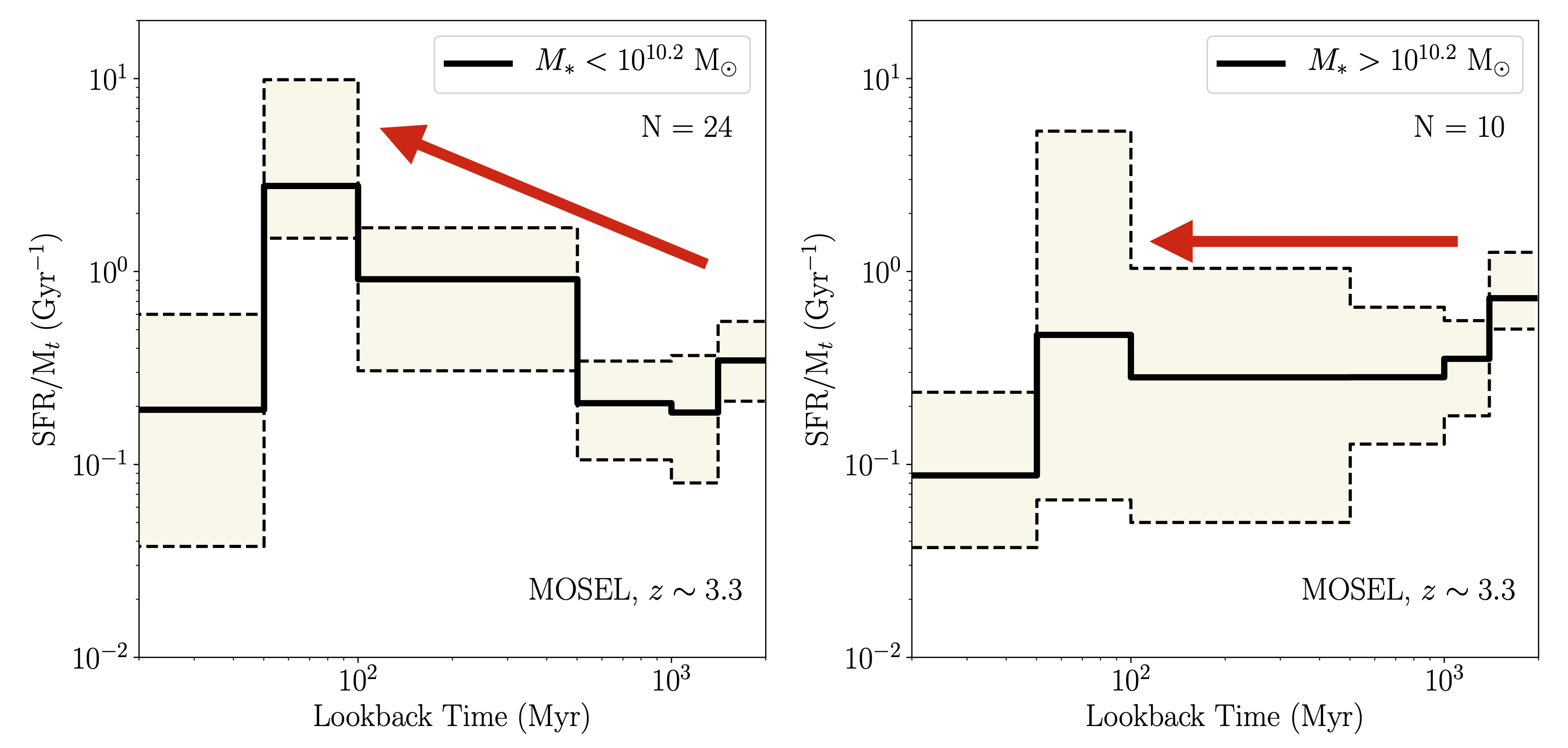}
	\includegraphics[scale=0.32, trim=0.0cm 0.0cm 0.0cm 0.0cm,clip=true]{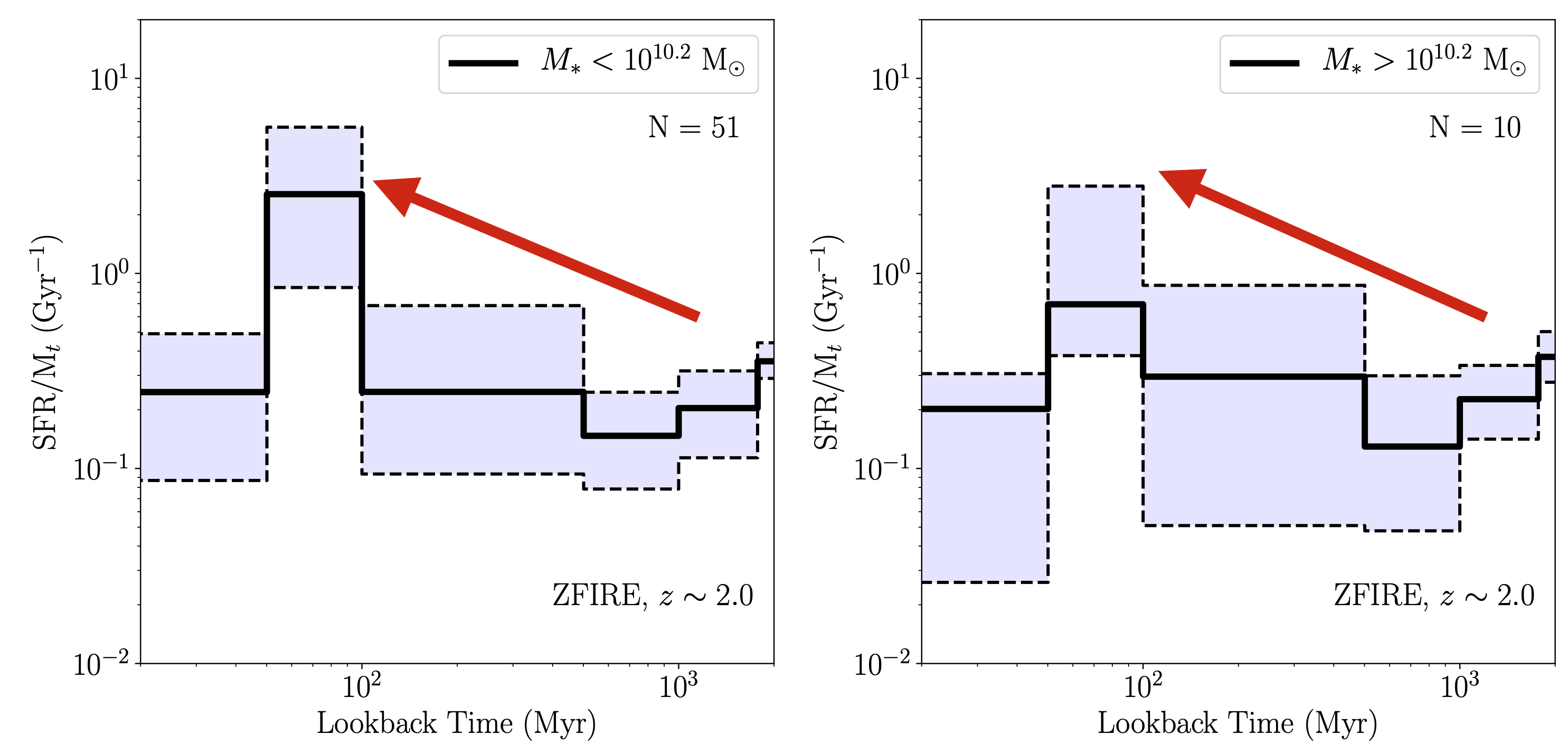}
	\caption{Star formation history of the low (\logmstar\,$< 10.2$; left) and high mass galaxy (\logmstar\,$> 10.2$; right) samples in the \mosel\ (top) and \zfire\ (bottom) surveys using \prospector\ \citep{Leja2017, Leja2019}. The  solid black line and  the shaded region in each panel represent the median and $16-84$ percentile regions, respectively. The red arrow in each panel shows the qualitative slope of SFHs in the respective sample. The massive \mosel\ galaxies have flat/declining SFHs in contrast to the SFHs in other samples.  }
	\label{fig:sfh_pros}
\end{figure*}

\subsection{Star formation histories of \mosel\ galaxies}\label{sec:sfhs}

The kinematic state of the gas is modulated by the star formation history (SFH) of the galaxy and the gas inflows/outflows. We use a python-based spectral energy distribution (SED) fitting code \prospector\ to recover the SFHs of \mosel\ galaxies \citep{Leja2017, Leja2019}. The extensive \zfourge\ photometry \cite{Straatman2016} provide us with fluxes in nearly 30 photometric bands for \mosel\ galaxies. 

Figure \ref{fig:sfh_pros} shows the recovered SFHs of the massive and low mass galaxies of our \mosel\ sample in comparison with the \zfire\ sample \cite{Nanayakkara2016}. We have normalized the SFR with the recovered stellar mass of each galaxy. The stellar mass estimates from \prospector\ are nearly $0.5$\,dex higher than the stellar mass estimated using FAST  \citep{Kriek2009} in the \zfourge\ survey, similar to the \cite{Cohn2018} observation. The stellar mass difference  between \prospector\ and FAST is due to the older stellar populations inferred by the non-parametric SFHs. Parametric SFH fit would be biased towards the younger stellar ages to explain the UV luminosity of galaxies.  Thus, the contribution of the older stellar population to the total mass will not be correctly constrained \citep{Leja2019a}. For consistency, we use the stellar mass estimates from FAST to separate galaxies into the two mass bins. 

The \mosel\ and \zfire\ samples are derived from the \zfourge\ surveys, allowing a consistent measurement of SFHs for both samples. We again separate the \zfire\ and \mosel\ galaxies into two mass bins at \logmstar\,$=10.2$.    To estimate the median and scatter in SFHs for each sample, we generate 1000 samples for each galaxy using the distribution of the posterior for each parameter. For each galaxy sample, we combine all randomly generated sample in each time bin and calculate $\rm 16^{th},\ 50^{th},\ {and}\ 84^{th}$\,percentile. The median and scatter in sSFR in the 6th time bin ($\sim 1.4-1.9\,$Gyr) is calculated without bootstrapping because it is not an independent variable in \prospector. 

Massive galaxies (\logmstar\,$>10.2$) in our \mosel\ sample show either a constant or declining star formation histories, whereas low mass galaxies (\logmstar\,$< 10.2$) have rising SFHs till 50 Myr ago. In contrast, there is no significant difference in SFHs of the low and high mass galaxies at $z\sim2$ from \zfire\ observations.  Massive galaxies at $z> 3$ only assemble $\sim 30$\% of their stellar mass in the past  500 Myr, whereas galaxies of similar stellar masses at $z\sim 2.0$  assemble more than $\sim 45$\% of their stellar mass in the past 500 Myr.  The low mass galaxies in the both \mosel\ and \zfire\ surveys assemble $\sim 65$\% of their stellar mass in the past 500 Myr.  Figure \ref{fig:sfh_pros} shows that massive galaxies at $z>3$ have nearly flat median SFHs, in contrast to the rising median SFHs of massive galaxies at $z=2$. We cannot derive statistically significant conclusions about the difference in the SFHs because of our limited sample size and large uncertainties.

Nearly constant SFHs of the massive galaxies (\logmstar\,$>10.2$) at $z> 3.0$ suggests their  relatively quiet evolution without a sudden influx of gas or mergers. We find that sSFRs drops in the 0-50\,Myr time bin irrespective of the sample, probably because non-parametric SFHs are better determining older stellar populations with age $>100\,$Myr \citep{Leja2019a}. A larger sample of galaxies and better photometric sampling in infrared bands for galaxies at $z>2.0$ is required to improve constraints on SFHs.

\begin{figure*}
	\centering
	\tiny
	\includegraphics[scale=0.6, trim=0.0cm 0.9cm 0.0cm 0.5cm,clip=true]{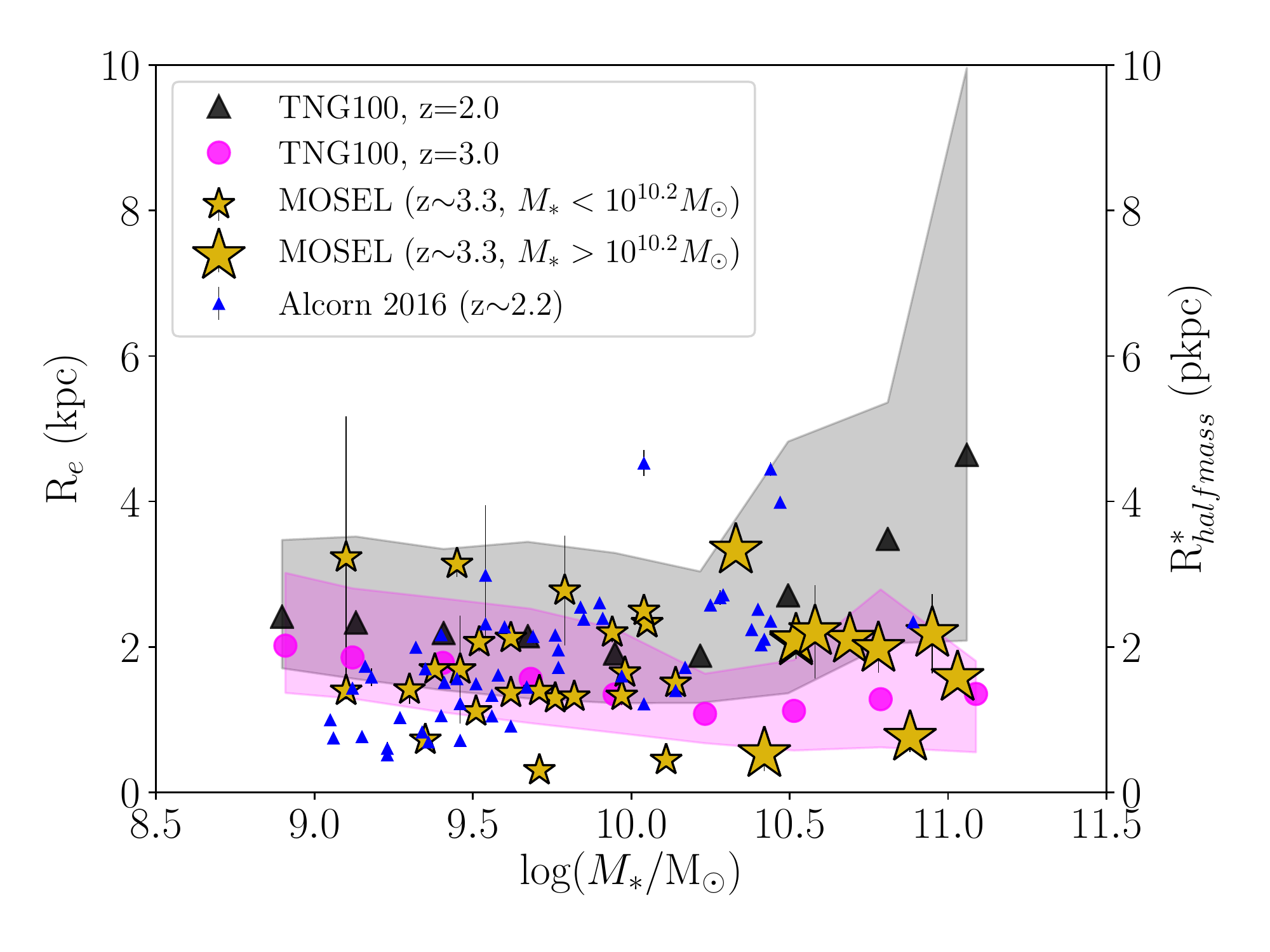}
	\caption{ Comparison between HST-F160W effective radii ($R_e$) from observations and $R^*_{halfmass}$ from the TNG100 simulation as a function of the stellar mass. The large and small golden stars correspond to the \mosel\ galaxies with \logmstar$>10.2$ and \logmstar$<10.2$ respectively.  The pink circles and shaded region represent the 50$^{th}$, 16$^{th}$ and 84$^{th}$ percentile in the $R^*_{halfmass}$  from TNG100 at $z=3.0$. Similarly, the black triangles and gray shaded region represent the $R^*_{halfmass}$ at $z=2.0$ in TNG100. }
	\label{fig:r80}
\end{figure*}

\begin{figure*}
	\centering
	\tiny
	\includegraphics[scale=0.25, trim=0.0cm 0.0cm 0.0cm 0.0cm,clip=true]{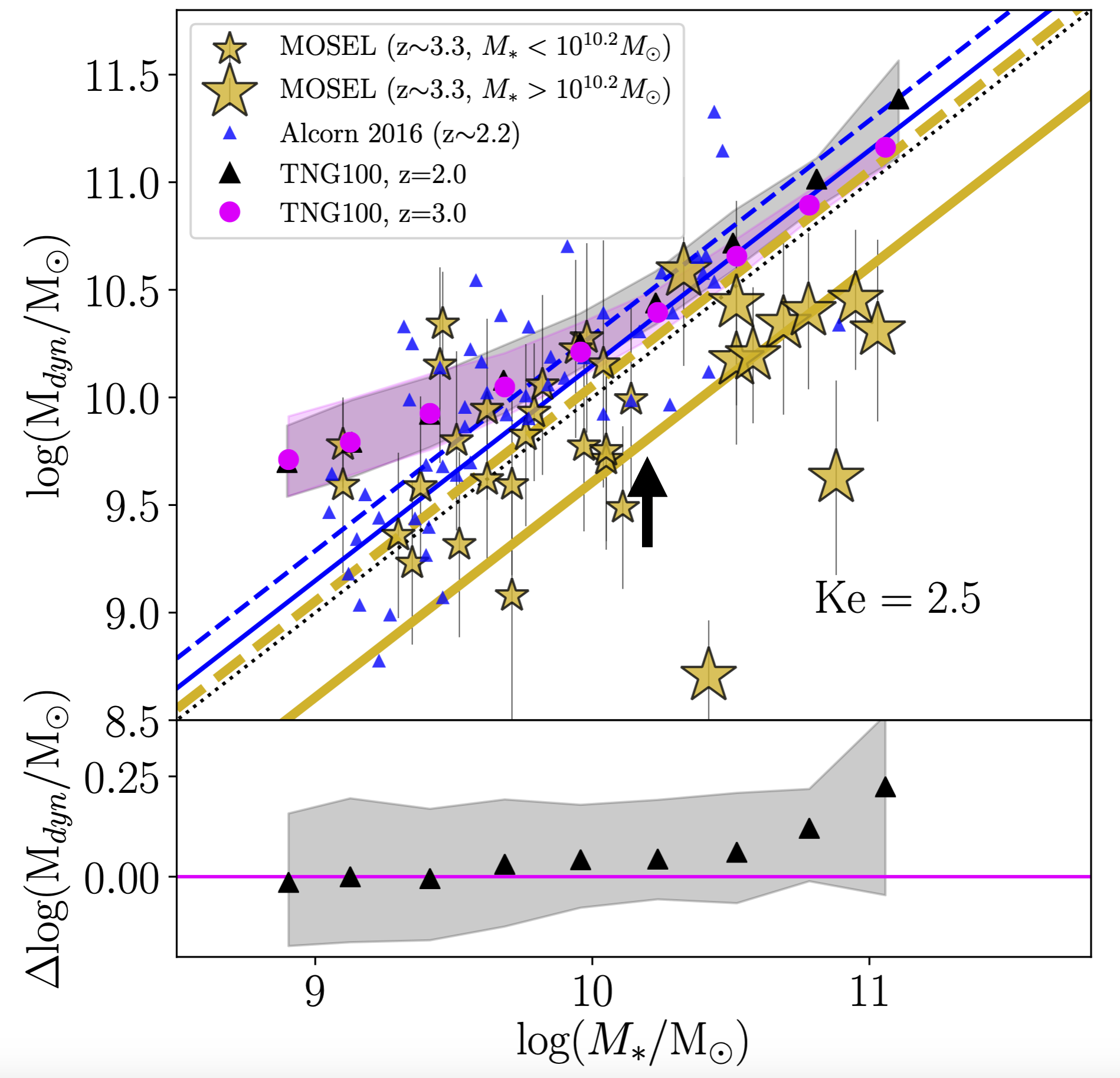}
	\includegraphics[scale=0.49, trim=0.0cm 0.0cm 0.0cm 0.0cm,clip=true]{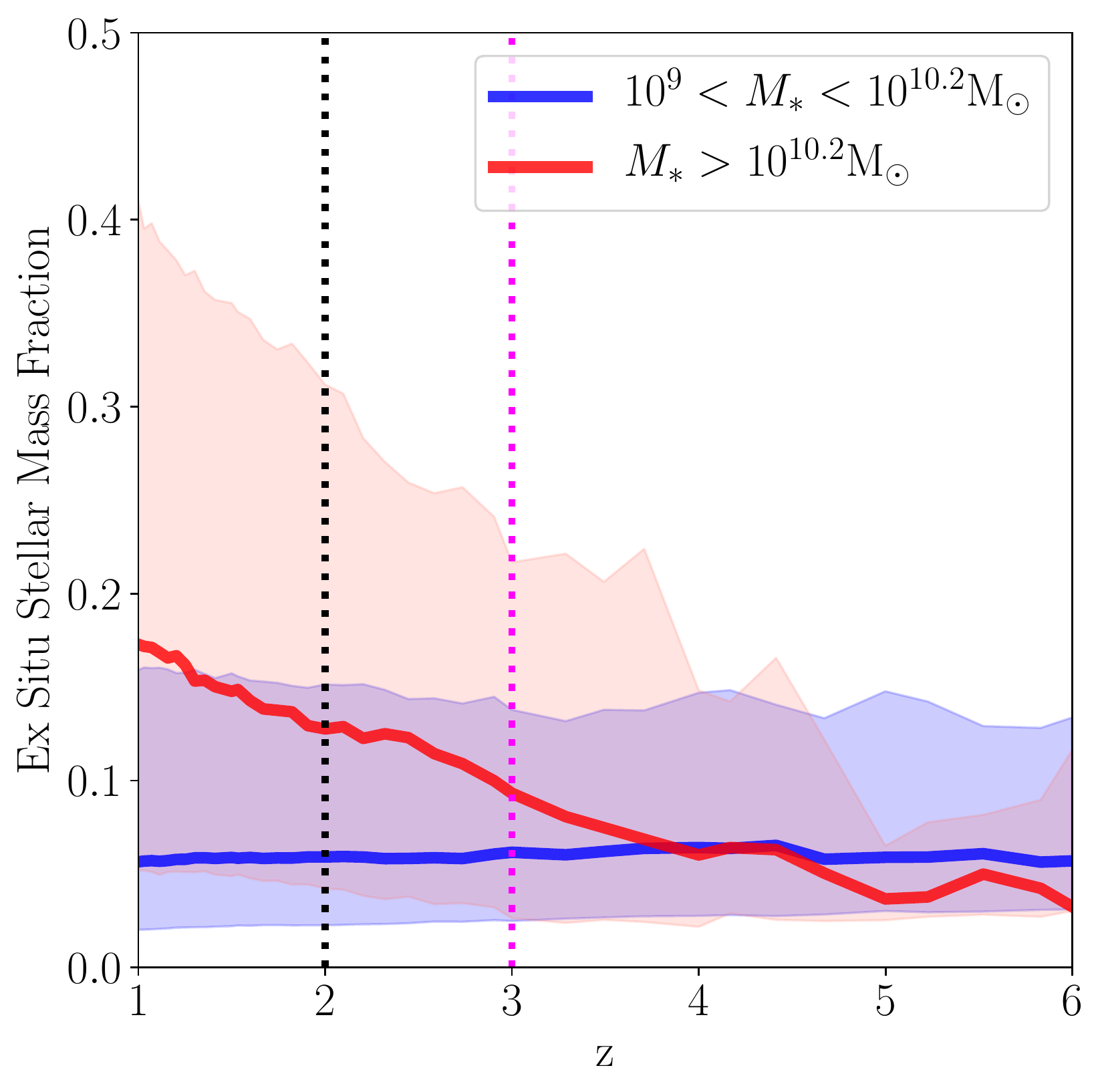}
	\caption{{\it Left panel:} Comparison between the dynamical to stellar mass relation in observations and simulations. The color scheme and symbols are the same as Figure \ref{fig:r80}.  The black dotted line is the one-to-one line.  The colored dashed and solid lines represent the average offset in the dynamical mass of low and high mass (split at \logmstar$=10.2$, pointed by the black arrow) galaxies in the respective samples from the one-to-one line.  The bottom panel shows the change in the dynamical mass at a fixed stellar mass between $z=2-3$, showing the $\sim0.1$\,dex increase in the dynamical mass at $z=2$ for massive galaxies.  {\it Right panel:} The  \exsitu\ stellar mass fraction (accreted) with respect to redshift for low mass ($9.0<$\,\logmstar\,$<10.2$, blue) and high mass sample (\logmstar\,$>10.2$, red) in the TNG100 simulation. The solid line  and shaded region represent the median and, 16$^{th}$, and 84$^{th}$ percentiles, respectively. The pink and black dotted lines mark the redshift snapshots of $z=3$ and $z=2$ respectively. The \exsitu\ stellar mass fraction increases sharply below $z<3.5$ for massive galaxies. }
	\label{fig:illustris_dm}
\end{figure*}

\section{Mass assembly in cosmological simulations }\label{sec:simulations}

Using slit-based spectroscopic observation of galaxies at $z\sim3.0$ in the \mosel\ survey, we find that massive galaxies (\logmstar\,$> 10.2$) have $56\,\pm\,21$\,km/s lower integrated velocity dispersion compared to galaxies in similar stellar mass range at $z\sim 2.0$  (Figure \ref{fig:st_mass_vel}).  We speculate that a lower dynamical mass for massive galaxies at $z=3$ compared to galaxies at $z=2$ could be responsible for their low velocity dispersion. 

We use a cosmological hydrodynamical simulation, IllustrisTNG \citep{Pillepich2017a, Nelson2017, Springel2017,  Marinacci2017, Naiman2017} to understand the evolution of the dynamical mass of galaxies. We use the $\sim$(100 Mpc)$^3$ volume (TNG100) simulation, because it has sufficient volume to produce a statistically significant sample of massive galaxies at $z>3$ (as opposed to TNG50).  TNG100 also has sufficient numerical resolution to reliably constrain the  properties of galaxies $M_* \sim 10^{9}$\msun. 

TNG100 has a baryonic mass resolution of $m_b = 9.4 \times 10^5$/h, where $m_b$ is the baryonic mass per particle. Selecting  galaxies from TNG100 at \logmstar\,$ > 9$ results in at least 1000 stellar particles per galaxy minimizing the numerical uncertainties.  To identify the progenitors of each selected galaxy, we track them back in time using the merger tree catalogs generated using the \cite{Rodriguez-Gomez2015} technique. An additional cut of SFR\,$>\,0$ is imposed while selecting galaxies at any redshift epoch because we aim to compare with kinematic measurements via emission lines that are intrinsically biased towards SFGs.

\subsection{Disk-size comparison between observations and simulations}

 To compare the dynamical mass between simulations and observations, we use the total mass enclosed within a stellar-half mass radius ($R^*_{halfmass}$) for simulated galaxies. Similar to \cite{Genel2017}, $R^*_{halfmass}$ is defined as the three-dimensional (3D) radius enclosing 50\% mass of all evolving stellar particles (stars plus stellar remnants) assigned to the galaxy by the SUBFIND algorithm. \cite{Genel2017} show that the  $R^*_{halfmass}$ is consistent within 1-sigma scatter to the 2D projected sizes in r-band across all stellar masses for both main-sequence and quenched galaxies.   Although, the 3D $R^*_{halfmass}$ is nearly $0.1-0.2$\,dex higher than the two-dimensional  half-light radii for simulated galaxies with \logmstar$<10.0$.

Figure \ref{fig:r80} shows the relation between $R^*_{halfmass}$ and stellar mass for the TNG100 galaxies across two redshift snapshots in comparison to the stellar mass-size relation of our \mosel\ galaxies. In simulations, the relation between $R^*_{halfmass}$ and the stellar mass remains consistent across $z=2-3$ within 1-sigma scatter. We do find an increased scatter in the $R^*_{halfmass}$ at the massive end at $z=2$ in simulations. 

Similar to \cite{Genel2017}, we find that the effective radii of galaxies in observations are slightly smaller than the $R^*_{halfmass}$ in simulations across both redshift intervals  at the low mass end.   Inherent observational bias against the extended low-surface brightness region, projection effects,   and  uncertainties in the mass-to-light ratio, especially at high redshift,  might be responsible for  the discrepancy in the galaxy size between observations and simulations \citep{Bernardi2017, Genel2017}.

\subsection{Dynamical mass evolution}

 Figure \ref{fig:illustris_dm} shows a comparative evolution of the dynamical mass in observations and simulations. We remeasure  the dynamical mass of \mosel\ galaxies and the \cite{Alcorn2016} sample at $z\sim 2.0$ using equation \ref{eq:virial_theorm} but with a virial factor $Ke=2.5$, to estimate the enclosed dynamical mass within the effective radii \citep{Courteau2014}.  

In simulations, we define dynamical mass as the total mass (dark + baryonic matter) enclosed within $R^*_{halfmass}$.  We bin the data into ten bins of equal stellar mass. Due to the small number of massive galaxies in TNG100, we only select stellar mass bins that have more than five galaxies.  We find a consistent  relation between the dynamical mass versus the stellar mass relation of simulated galaxies across $z=2-3$, at least for \logmstar\,$<10$. We find a systematic upturn in the dynamical mass of simulated galaxies  at the massive end. The mean dynamical mass of simulated galaxies  with \logmstar\,$>10.0$ increases by roughly $0.1$\,dex between $z=2-3$.

  Observational measurements of dynamical mass from integrated spectra are riddled with unknowns such as mass to light ratio, projection effects, kinematic profiles, and S/N, making a direct comparison of dynamical mass between observations and simulations difficult. We find that the dynamical mass of simulated galaxies at $z=3$ is systematically $\sim$0.4\,dex higher than the massive galaxies in our \mosel\ sample.  The lower dynamical mass of massive \mosel\ galaxies compared to simulated galaxies can be due to our choice of a virial factor that is true for only disky-galaxies and would underestimate the dynamical mass of compact massive galaxies with high Sersic index \citep{Cappellari2006a, Courteau2014}.

 	The \cite{Alcorn2016} sample at $z\sim2.0$ only extends up to \logmstar\,$<10.5$, so we cannot compare the dynamical mass estimates of the massive \mosel\ galaxies with galaxies at $z=2$.  The  $R^*_{halfmass}$ for simulated galaxies is nearly two times larger than 2D-projected half-light radii for observed galaxies with \logmstar$<9.5$ (Figure \ref{fig:r80}), which might be responsible for $\sim$0.5\,dex higher dynamical mass of low mass simulated galaxies compared to the observations. 
 Within the limitation of our observational data, we do not find any systematic difference between the dynamical mass of the \mosel\ sample at $z>3.0$ and the \zfire\ sample from \cite{Alcorn2016} at $z\sim 2$. A larger sample of photometric and spectroscopic data between $z=2-4$ is required to observationally identify changes in the dynamical mass of galaxies between $z=2-4$.

\subsection{ {\it In situ} versus {\it ex situ} growth}

We use the \cite{Rodriguez-gomez2016} stellar assembly catalog to estimate the evolution of  \exsitu\ stellar mass fraction. \cite{Rodriguez-gomez2016} defines the \exsitu\ stellar mass fraction as the fractional amount of stellar mass for a galaxy that is contributed by the stars formed in other galaxies, which were subsequently accreted in the galaxy.  The \exsitu\ stellar mass fraction gives us a handle on the amount of stellar mass growth from accretion versus the {\it in situ} star formation. 

The right panel in Figure \ref{fig:illustris_dm} shows the evolution in the \exsitu\ stellar mass fraction with redshift. At each redshift epoch, we select simulated galaxies with non zero SFRs and split them into two stellar mass bins at \logmstar\,$=10.2$ to match with our observations. In the low stellar mass bin, we only select galaxies with \logmstar\,$>9.0$ to minimize numerical uncertainties.  We calculate the $\rm 50^{th},\ 16^{th},\  and\ 84^{th}$ percentiles in the \exsitu\ stellar mass fraction for the two stellar mass bins. 

Figure \ref{fig:illustris_dm} clearly shows a systematic increase in the \exsitu\ stellar mass fraction of massive simulated galaxies.  The low mass galaxies accrete roughly 6\% of their stellar mass from other galaxies, and the fraction remains unchanged until $z=1.0$. In contrast, massive simulated galaxies accrete $\sim 6.8\%$ of their stellar mass from other galaxies until $z\sim 3.5$ that subsequently increases rapidly. The median \exsitu\ stellar mass fraction for massive galaxies changes from $\sim$ 9\% at $z=3$ to $\sim$ 13\% at $z=2$ and reaches to about 17\% by  $z=1$. The increased scatter in the \exsitu\ stellar mass fraction for massive galaxies towards lower redshift might be driven by the absolute increase in the total number of massive galaxies at low redshifts. 

Our choice of the stellar mass cut is nearly equal to $M^*$ for the SFG population at $z=2.5-3$, where $M^*$ is the turn-over mass in the stellar mass function \citep{Davidzon2017}.  Changing the stellar mass cut-off to \logmstar\,$=\,10.0$ pushes the redshift at which \exsitu\ stellar mass fraction starts to rise to slightly lower redshifts without significantly altering the systematic trend. Our result is consistent with the \cite{Rodriguez-gomez2016} analysis, who use the original Illustris simulation to find that the transition from {\it in situ} to \exsitu\ stellar mass growth occurs only for the most massive galaxies at $z\approx1.0$.  

We suspect that the stellar mass growth via \exsitu\ processes might be responsible for the increase in the integrated velocity dispersion of massive galaxies between $z=3.0$ to $z=2.0$ (see Section \ref{sec:discussion} for further discussion).

\section{Discussion}\label{sec:discussion}
By measuring the \oiii\ emission line profile from MOSFIRE observations, we find that massive galaxies (\logmstar\,$> 10.2$) at $z>3$ have nearly $56\,\pm\,21$\,km/s lower integrated velocity dispersion than similar stellar mass galaxies at $z\sim2$ (Figure \ref{fig:st_mass_vel}). We also find that massive galaxies at $z>3.0$ have either flat or declining SFHs, in contrast,  galaxies of similar stellar mass at $z\sim 2.0$ have slightly rising SFHs (Figure \ref{fig:sfh_pros}). The integrated velocity dispersion represents a combination of the rotation velocity and intrinsic velocity dispersion of galaxies, thus giving us a handle on both the kinematic properties of  gas and the total mass budget of galaxies. In the following subsections, we try to disentangle the evolution of the intrinsic velocity dispersion from the mass assembly history of galaxies to explain our observations. 

\subsection{Kinematics  of gas and SFHs}

Large surveys such as \kmos have shown a significant evolution in the  kinematics of ionized gas between $z=1-3$  \citep{Wisnioski2015}.  Both local and high redshift galaxies show a correlation between the intrinsic velocity dispersion of gas and their star formation rate, albeit with a significant secondary dependence on other galaxy properties such as gas fraction \citep{Krumholz2016}. Internal secular processes such as evolving gas reservoirs,  higher star formation rate and gravitational instabilities introduced by the gas accretion and outflows, drive the higher intrinsic velocity dispersion of high redshift galaxies  \citep{Newman2013a, Krumholz2016, Wiseman2017, Ubler2018, Zabl2019, Martin}. 

The cosmic star formation density peaks at $z\sim2.0$ \citep{Madau2014}. The declining cosmic SFR density at $z>2$ could lead to a decline in the intrinsic velocity dispersion of galaxies at $z>2.0$.  \cite{Saintonge2013} also find evidence of a flattening or decrease in the cold gas fraction for galaxies at $z>2.8$, which could translate into lower intrinsic velocity dispersion.  Current observational studies do not  show any conclusive evidence of a decline in the intrinsic velocity dispersion of massive galaxies between $z>2.0$ \citep{Turner2017, Ubler2019}. 

Most integral field spectroscopic observations have small numbers  of galaxies at $z>3$ especially at \logmstar\,$>10.0$ \citep{Gnerucci2011, Wisnioski2015, Girard2018}.  With a sample of 11 galaxies at $z>3$ in the \logmstar\,$=9.0-11.0$, \cite{Gnerucci2011} find the intrinsic velocity dispersion of galaxies is $\sim 60$\,km/s. By combining data from various observations between $z=1-3.5$, \cite{Wisnioski2015} find that the intrinsic velocity dispersion of galaxies with \logmstar$\,>10.5$ increases from $\sim50$\,km/s at $z=2$ to $\sim70$\,km/s at $z\sim3$.  Similarly, \cite{Turner2017} find that galaxies at $z>3$ have nearly 70\,km/s intrinsic velocity dispersion. However, a monotonic rise in the intrinsic velocity dispersion with redshift is opposite to our observation of a lower integrated velocity dispersion for galaxies at $z>3$ compared to galaxies of similar stellar masses at $z\sim 2.0$.

\cite{Girard2018} analyze the kinematics of 24 gravitationally lensed galaxies at $z=1.4-3.5$ as a function of stellar mass and redshift.  They find no significant evolution in the intrinsic velocity dispersion of low mass galaxies (\logmstar$\,<\,10$) between $z\sim 3.0$ to $z\sim2.0$, similar to our \sigmaint\ measurements for the low mass galaxy sample (Figure \ref{fig:st_mass_vel}).  By separating galaxies into two stellar mass bins at \logmstar$\,=10.2$, \cite{Girard2018} find $\sim15$ km/s lower intrinsic velocity dispersion for massive galaxies compared to the low mass galaxies. They suspect irregular sampling might be responsible because the average redshift of their low mass sample is  $z\sim3.1$, in contrast, the average redshift of high mass sample is $z\sim2.4$. 

 In lieu of the lack of any conclusive evidence that the intrinsic velocity dispersion of massive galaxies declines or steadily rises between $z=2-4$, we cannot rule out a lower intrinsic velocity dispersion of massive galaxies at $z>3.0$ compared to galaxies of similar stellar masses at $z\sim2$. In the following subsection, we discuss if a difference in the mass assembly history can explain our observations.

\subsection{Mass assembly history}
 
 Kinematic properties of gas and stars are a powerful tool to understand the relative contribution of various physical processes such as monolithic collapse of gas \citep{Eggen1962, Searle1978}, smooth gas accretion \citep{Fall1980}, and galaxy-galaxy mergers \citep{White1978}   to the mass assembly history of galaxies. Observational studies find an increasing role of the baryonic component to the total mass budget of galaxies at higher redshifts \citep{ ForsterSchreiber2009, Gnerucci2011, Simons2016, Straatman2017, Glazebrook2017,   Price2019}. \cite{Ubler2017} find between $z=0.9-2.3$ the zeropoint of the stellar mass Tully-Fisher relation (TFR)  does not change but the baryonic TFR decrease significantly. The higher baryonic content  of high redshift galaxies at a fixed stellar mass is driven by the rising gas fraction of galaxies with redshift \citep{Saintonge2013, Tacconi2018}.

\cite{Gnerucci2011} find that the zeropoint of TFR is lower by $0.88$\,dex  for galaxies at $z>3.0$ compared to $z\sim 2.0$, albeit with a significant scatter.  \cite{Price2019} also find that the dark matter fraction of galaxies decreases with redshift until $z\sim 3.5$. The $z>3.0$ sample of \cite{Price2019} extends only till \logmstar\,$<10.5$ compared to our \logmstar\,$\sim 11.0$.   Within the limited sample and scatter, our massive \mosel\ galaxies have $\sim 0.4$ dex lower dynamical mass compared to the same stellar mass galaxies at $z\sim2$ in observations (Figure \ref{fig:dynamical_mass}). 

In IllustrisTNG simulations, we find a $0.1$\,dex increase in the dynamical mass of massive simulated galaxies at a fixed stellar mass between $z=2-3$ (Figure \ref{fig:illustris_dm}).  Observational estimates of the dynamical mass roughly follow a similar relation to simulations, albeit with a larger scatter. We suspect that not accounting for, e.g. the likely higher Sersic index and compact structure, of our massive MOSEL galaxies may account for the $\sim0.4\,$dex lower dynamical mass compared to the same stellar mass galaxies from the  IllustrisTNG simulation.

Our observation of a lower integrated velocity dispersion of massive galaxies at $z=3$ compared to  galaxies of similar stellar masses at $z\sim2$ could be probing the changing rotation velocity profile of massive galaxies due to the evolving baryonic fraction.  We note  that the resolved kinematic observations of massive galaxies at $z>3$ are required to confirm the changing rotation profile of galaxies. \cite{Lang2016} and \cite{Genzel2017}  find a turnover in the rotation velocity of the gas  in galaxies at $z\sim2.0$. They speculate that a lower concentration of dark matter in the inner galactic disks, resulting from the ongoing dark matter assembly and  asymmetric drift pressure is   responsible for the turnover in the rotation velocity profile.  \cite{Teklu2018}, using {\it Magneticum Pathfinder} simulations \citep{Beck2016},  show that even after including the asymmetric drift pressure support for cold gas, almost 50\% of their galaxies exhibit a turnover in their rotation curves, indicative of the low dark matter fraction in high-redshift galaxies.  However, observations of resolved rotation profiles are susceptible to the variable spatial resolution and size evolution of galactic disks  \citep{Tiley2019}.

 We find that in the IllustrisTNG simulation,  the $0.1$\,dex rise in the dynamical mass to the stellar mass fraction of massive galaxies at $z=2.0$ is coupled to a rise in the \exsitu\ stellar mass fraction.  The \exsitu\ stellar mass fraction of massive galaxies (\logmstar\,$>10.2$) increases by a factor of  two between $2<z<3.5$ (Figure \ref{fig:illustris_dm}). In contrast, the \exsitu\ stellar mass fraction of low mass galaxies remains nearly constant. 
	
The rising contribution of \exsitu\ processes such as mini and minor mergers can be responsible for the nearly 0.1\,dex higher dynamical mass of massive galaxies at $z=2$ compared to $z=3$ in simulations \citep[Figure \ref{sec:dy_mass}; ][]{Hilz2013}. The \exsitu\ processes through the accretion of gas and stars can drive significant turbulence and gravitational instabilities in the galactic disks \citep{Genel2012, Mandelker2014},   which in turn can result in a higher intrinsic velocity dispersion of galaxies \citep{Krumholz2018}.   

We speculate that observation of a higher integrated velocity dispersion of massive galaxies at  $z=2.0$ compared to galaxies of similar stellar masses at $z>3$ is probing the transition from the  {\it in situ} to \exsitu\ in the stellar mass assembly history of massive galaxies.  Rising SFHs of massive galaxies at $z=2.0$ also supports that massive galaxies at $z=2.0$ have acquired a fresh supply of gas in the past 500\,Myr (Figure \ref{fig:sfh_pros}).

\section{Summary}\label{sec:summary}

In this work, we combine near-infrared spectroscopic observations from MOSFIRE/Keck, deep \zfourge\ photometry, and  IllustrisTNG  simulations to analyze the mass assembly histories of galaxies at $z>3$. Our main results are:

\begin{enumerate}
	\item By measuring the \oiii\ emission profile of galaxies at $z\sim 3.3$, we find that galaxies with \logmstar\,$>10.2$ have $56\,\pm\,21\,$km/s lower integrated velocity dispersion compared to  galaxies of similar stellar masses at $z\sim2.0$ (Figure \ref{fig:st_mass_vel}). 
	
	\item We convert the integrated velocity dispersion into the dynamical mass of galaxies using virial theorem and find that massive galaxies at $z> 3$ have $\sim 0.4$\,dex lower dynamical mass compared to galaxies  of similar stellar masses at $z\sim 2$ (Figure \ref{fig:dynamical_mass}). 
	
	\item  We use \prospector\ to estimate star formation histories of galaxies from the \zfire\ and \mosel\ surveys, and find  that massive  galaxies at $z> 3$ have either flat or declining star formation histories till 50\,Myr. In contrast, similar stellar mass galaxies at $z\sim2$ show a slight peak in their SFH in the last 50\,Myr (Figure \ref{fig:sfh_pros}).  
	 
	 \item  Using IllustrisTNG simulations, we find a systematic $0.1$\,dex increase in the  dynamical to stellar mass ratio of massive simulated galaxies ( \logmstar\,$>10.0$) at $z=2$ compared to  $z=3$ galaxies (Figure \ref{fig:illustris_dm}).  
	 
	 \item By probing the stellar mass assembly histories of simulated galaxies, we find that a rapid rise in the \exsitu\ stellar mass fraction of massive galaxies (\logmstar\,$>10.2$) at $z<3.5$.  In contrast, the \exsitu\ stellar mass fraction of low mass sample remains constant across cosmic time (Figure \ref{fig:illustris_dm}).
\end{enumerate}

We speculate that the high integrated velocity dispersion and rising SFHs of  massive galaxies at $z\simeq2.0$ compared to  galaxies  of similar stellar masses at $z>3.0$  are driven by the rising contribution of \exsitu\ stellar mass to the total stellar mass growth  of massive galaxies. However, our conclusions are limited by the low signal-to-noise, limited sample size and heterogeneous stellar mass coverage of existing data. Large spectroscopic and photometric surveys of galaxies between $z=2-4$ with future facilities like GMT, ELT, MSE and LSST will provide sufficient samples and depth to test this hypothesis. 

\acknowledgments

The authors thank the referee for providing useful comments
and suggestions to improve the quality of the paper. K. Tran acknowledges support by the National Science Foundation under Grant Number 1410728. T.Y. acknowledges support from an ASTRO 3D fellowship. GGK acknowledges the support of the Australian Research Council through the Discovery Project DP170103470. Parts of this research were conducted by the Australian Research Council Centre of Excellence for All Sky Astrophysics in 3 Dimensions (ASTRO 3D), through project number CE170100013. TN acknowledge the Nederlandse Organisatie voor Wetenschappelijk Onderzoek (NWO) top grant TOP1.16.057.
The IllustrisTNG simulations and the ancillary runs were run on the HazelHen Cray XC40-system (project GCS-ILLU), Stampede supercomputer at TACC/XSEDE (allocation AST140063), at the Hydra and Draco supercomputers at the Max Planck Computing and Data Facility, and on the MIT/Harvard computing facilities supported by FAS and MIT MKI. The authors wish to recognize and acknowledge the very significant cultural role and reverence that the summit of Mauna Kea has always had within the indigenous Hawaiian community. 
We are most fortunate to have the opportunity to conduct observations from this mountain.

\bibliographystyle{aasjournal}

\appendix

\begin{figure*}
	\centering
	\tiny
	\includegraphics[scale=0.27, trim=0.0cm 0.0cm 0.0cm 0.0cm,clip=true]{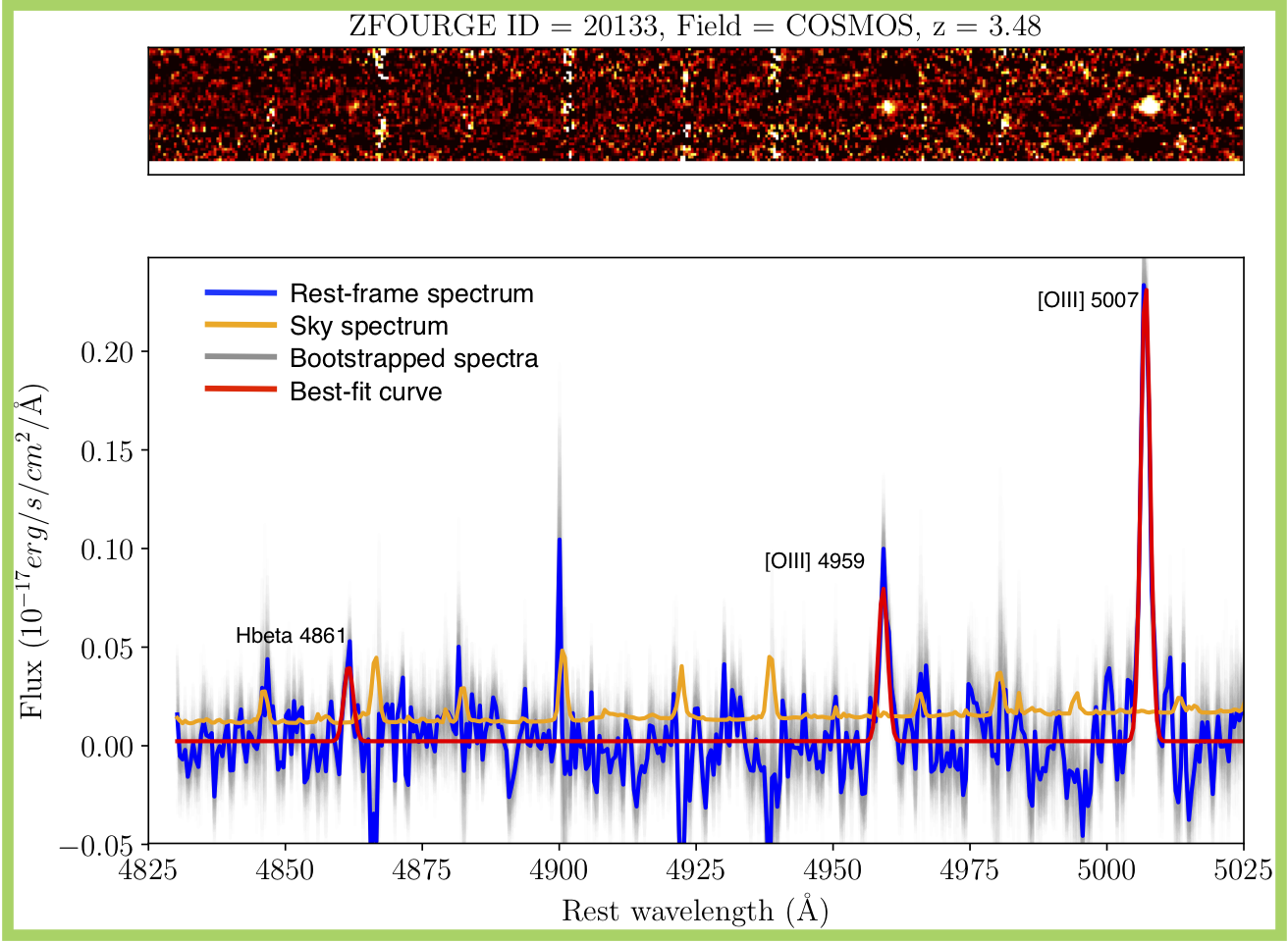}
	\includegraphics[scale=0.27, trim=0.0cm 0.0cm 0.0cm 0.0cm,clip=true]{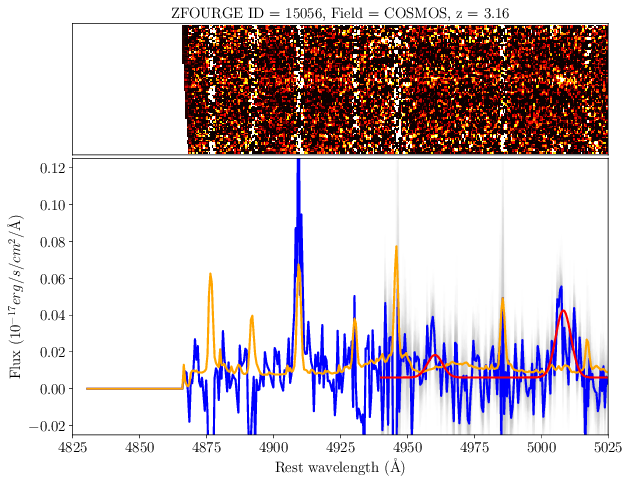}
	\includegraphics[scale=0.27, trim=0.0cm 0.0cm 0.0cm 0.0cm,clip=true]{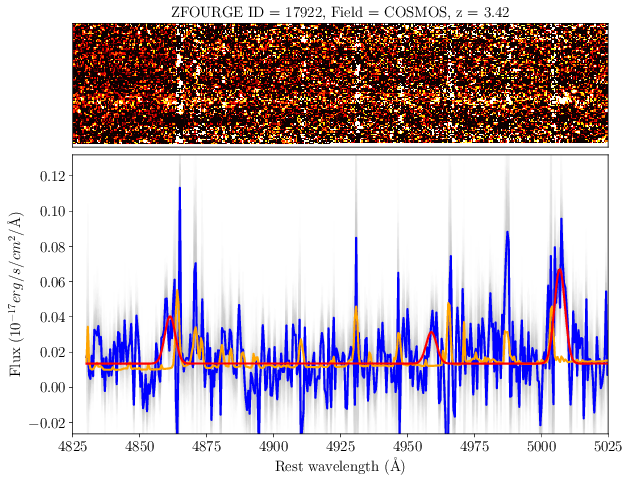}
	\includegraphics[scale=0.27, trim=0.0cm 0.0cm 0.0cm 0.0cm,clip=true]{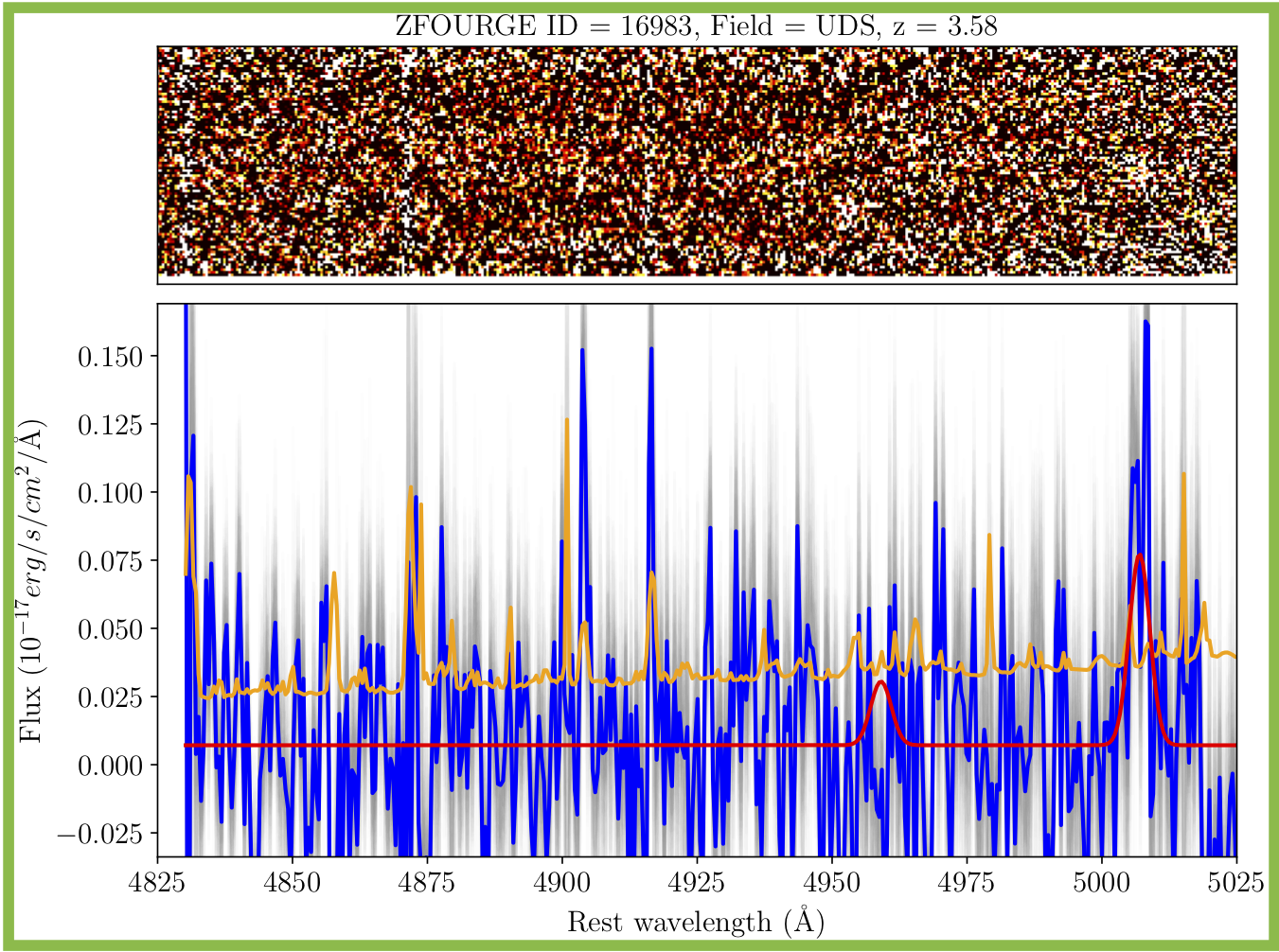}
	\includegraphics[scale=0.27, trim=0.0cm 0.0cm 0.0cm 0.0cm,clip=true]{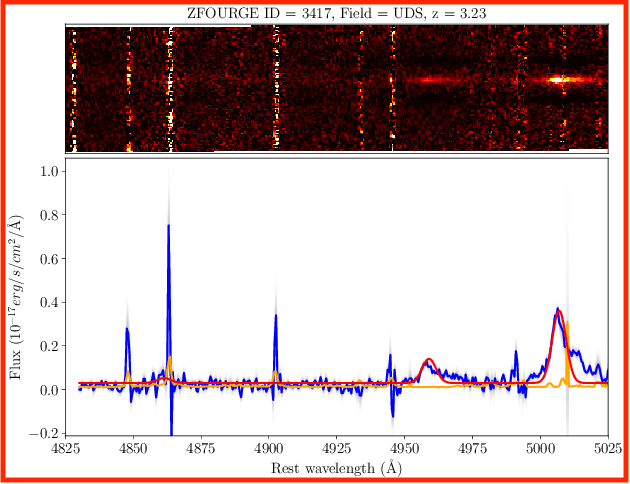}
	\includegraphics[scale=0.27, trim=0.0cm 0.0cm 0.0cm 0.0cm,clip=true]{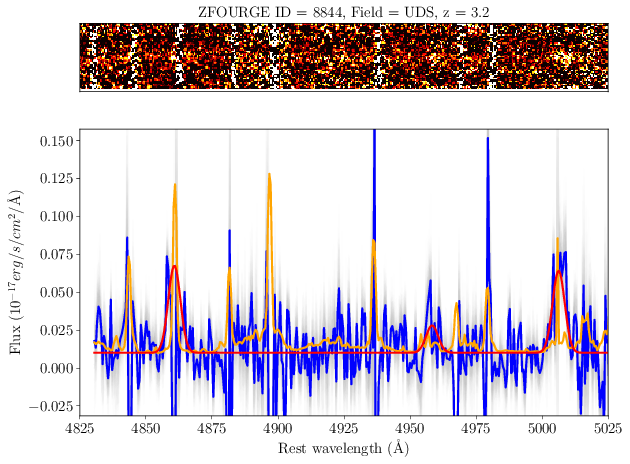}
	\includegraphics[scale=0.27, trim=0.0cm 0.0cm 0.0cm 0.0cm,clip=true]{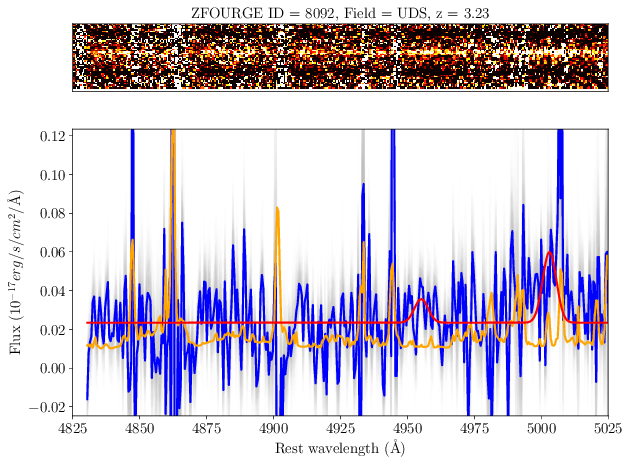}
	\includegraphics[scale=0.27, trim=0.0cm 0.0cm 0.0cm 0.0cm,clip=true]{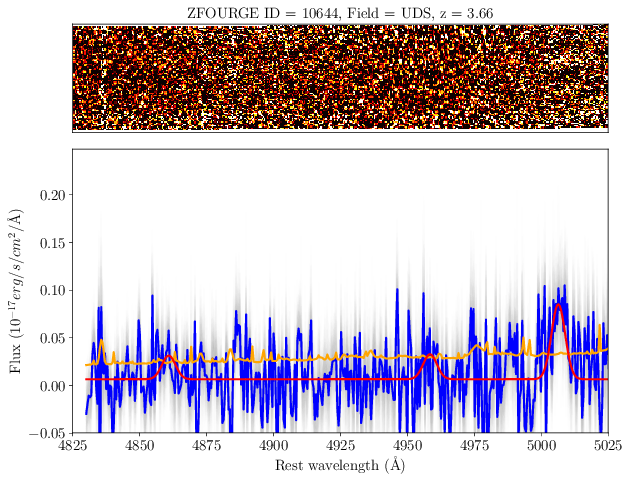}
	\includegraphics[scale=0.27, trim=0.0cm 0.0cm 0.0cm 0.0cm,clip=true]{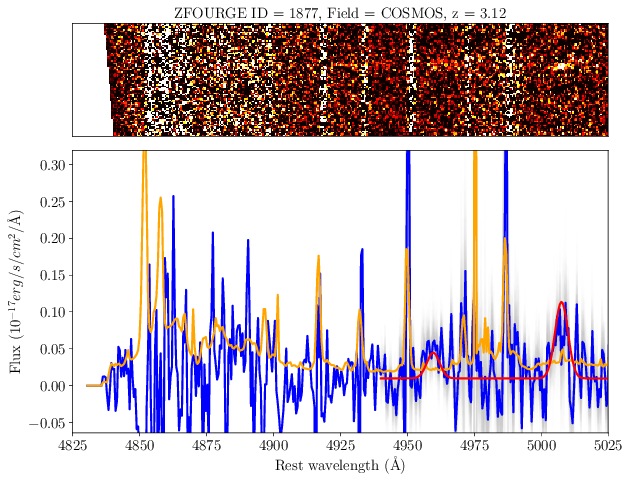}
	\includegraphics[scale=0.27, trim=0.0cm 0.0cm 0.0cm 0.0cm,clip=true]{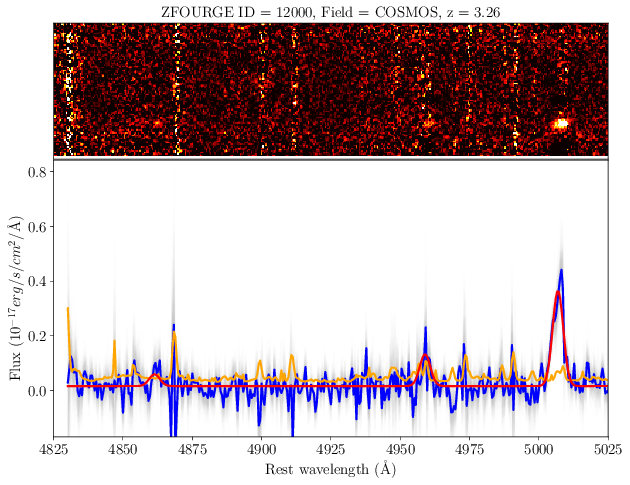}

	\caption{Same as Figure \ref{fig:spectra} but for all the 10 galaxies with \logmstar\,$>10.2$.  Spectra of two galaxies with unphysical dynamical masses are highlighted with green box. The spectrum highlighted with a red box corresponds to a possible AGN contaminant. }
	
	\label{fig:spectra_out}
\end{figure*}

\begin{figure*}
\centering
\tiny
\includegraphics[scale=0.30, trim=0.0cm 0.0cm 0.0cm 0.0cm,clip=true]{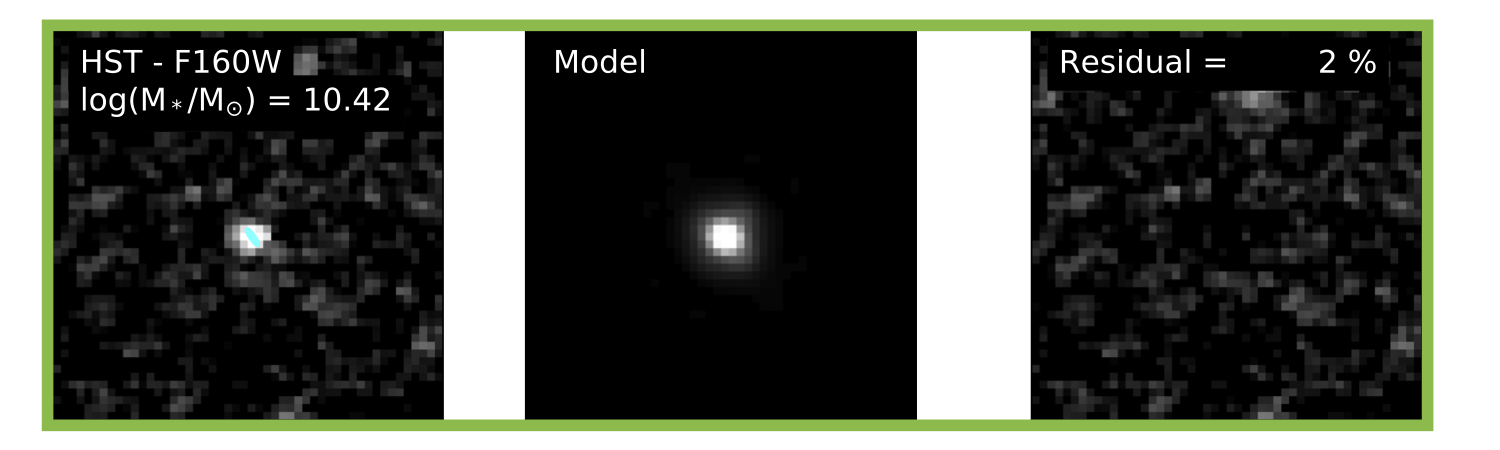}
\includegraphics[scale=0.30, trim=0.0cm 0.0cm 0.0cm 0.0cm,clip=true]{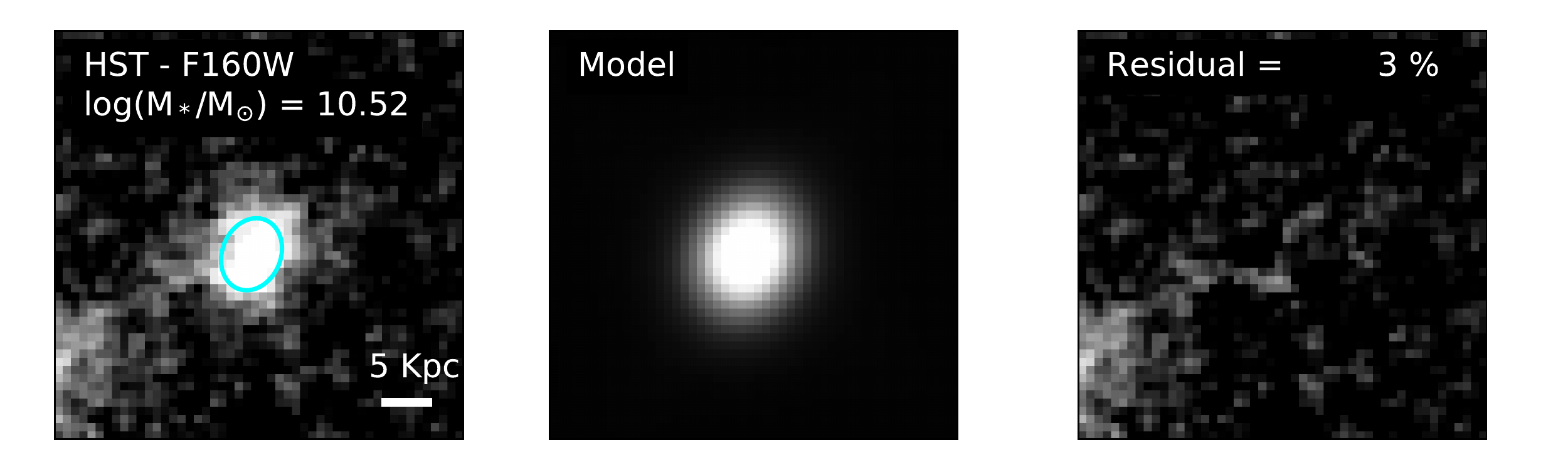}
\includegraphics[scale=0.30, trim=0.0cm 0.0cm 0.0cm 0.0cm,clip=true]{./galfit_out_11.pdf}
\includegraphics[scale=0.30, trim=0.0cm 0.0cm 0.0cm 0.0cm,clip=true]{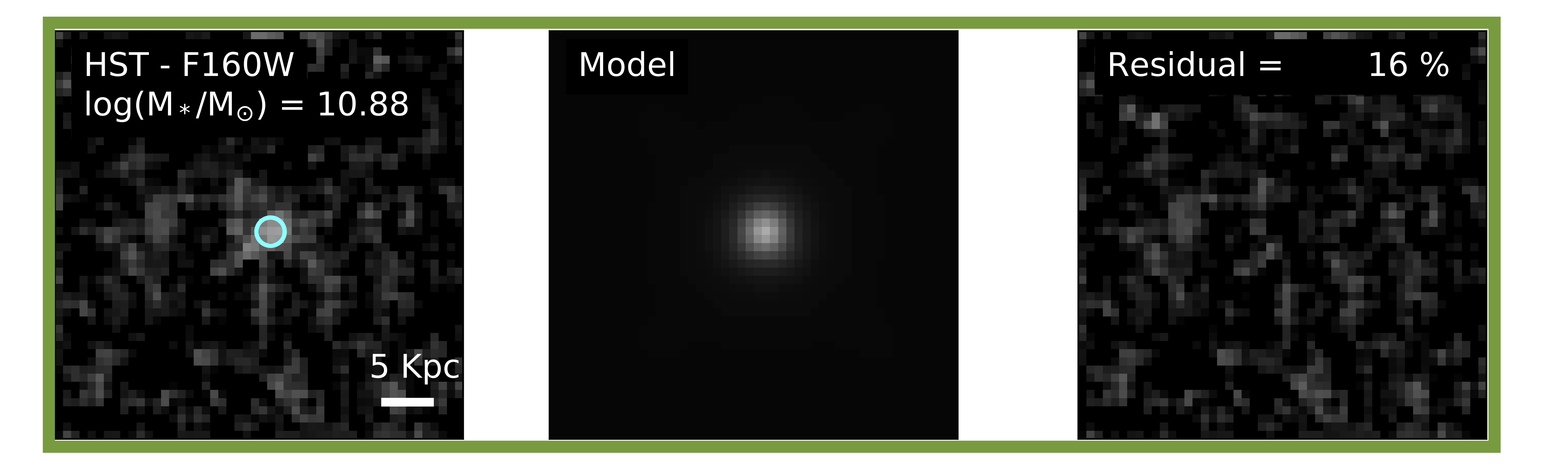}
\includegraphics[scale=0.30, trim=0.0cm 0.0cm 0.0cm 0.0cm,clip=true]{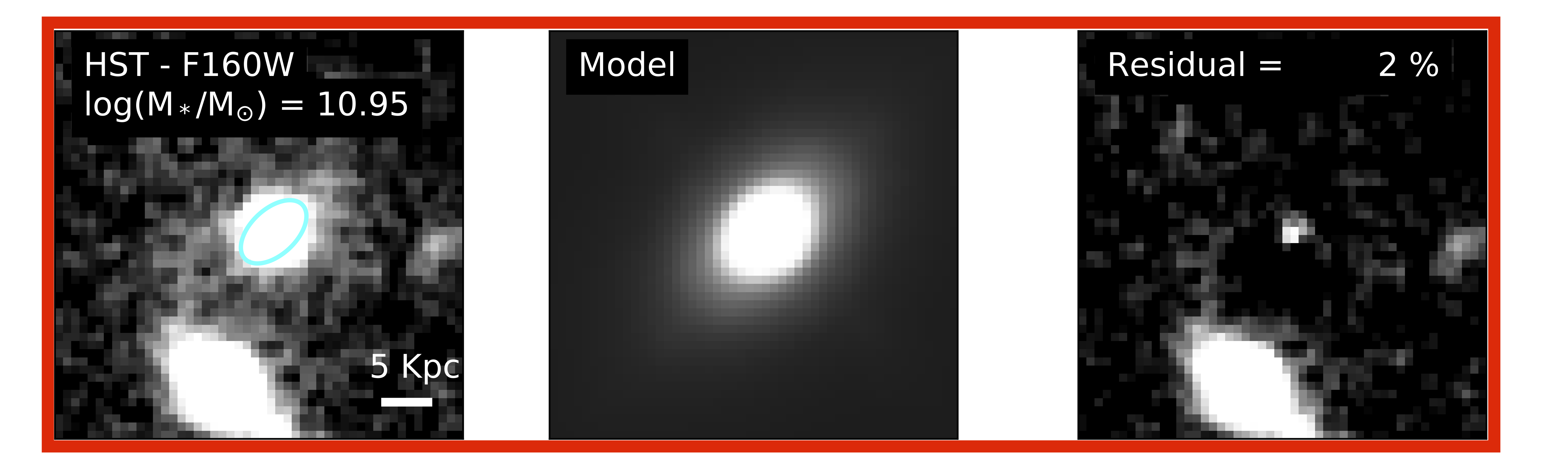}
\includegraphics[scale=0.30, trim=0.0cm 0.0cm 0.0cm 0.0cm,clip=true]{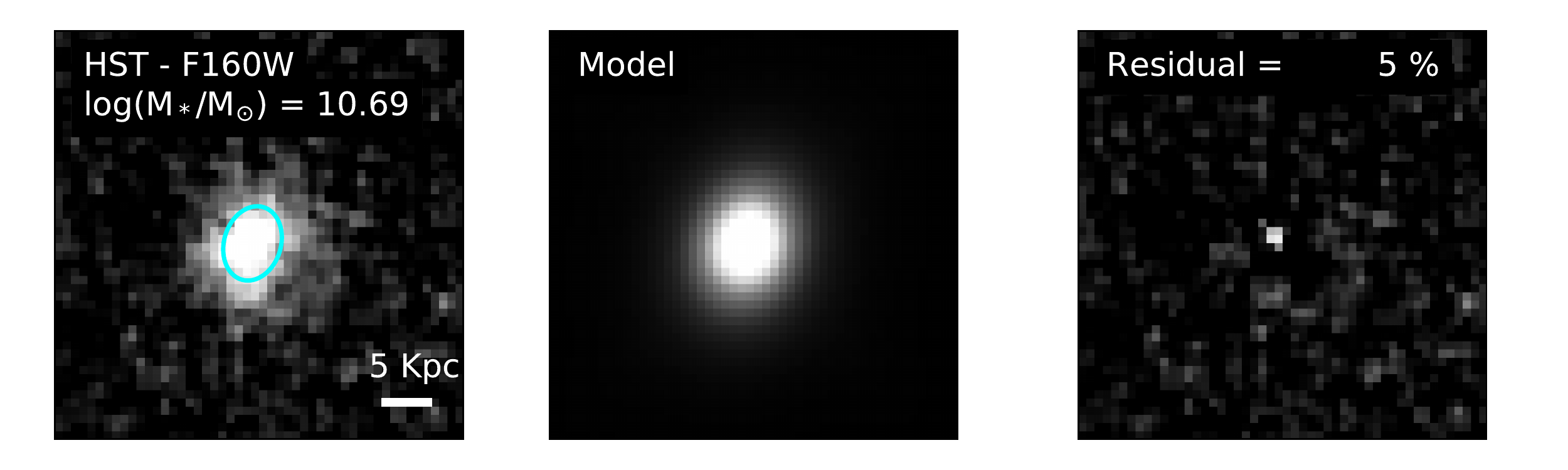}
\includegraphics[scale=0.30, trim=0.0cm 0.0cm 0.0cm 0.0cm,clip=true]{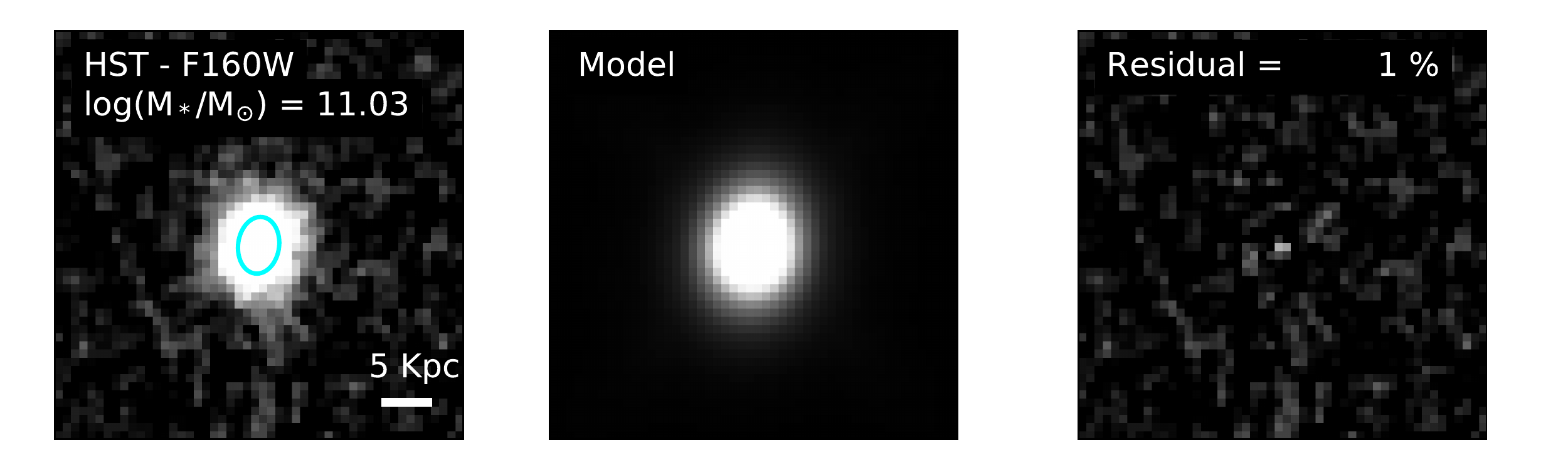}
\includegraphics[scale=0.30, trim=0.0cm 0.0cm 0.0cm 0.0cm,clip=true]{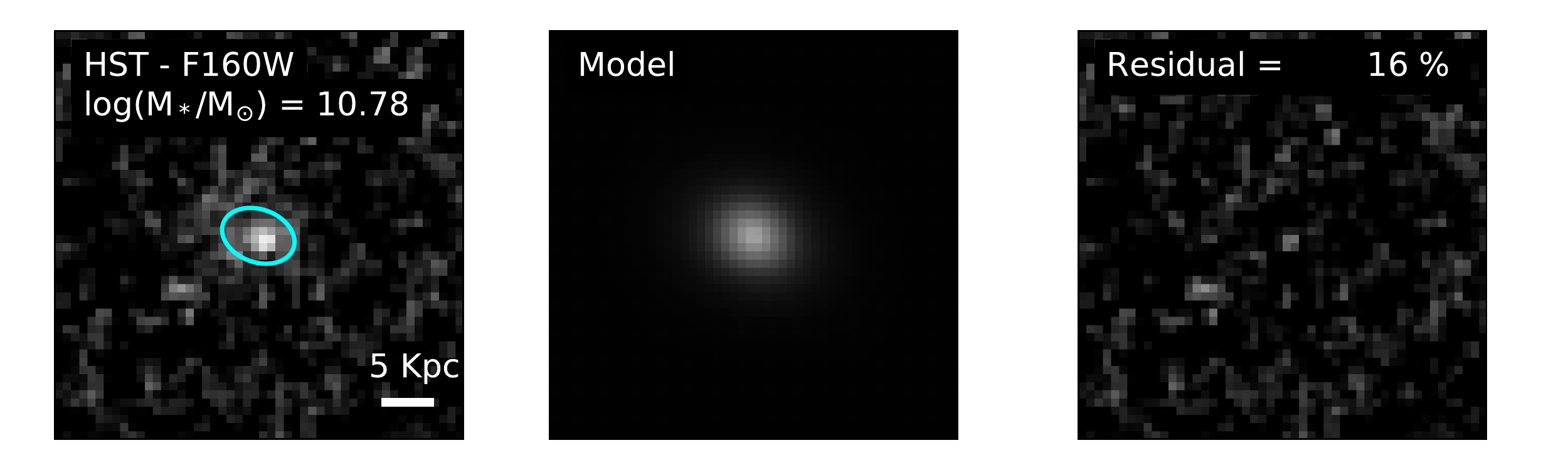}
\includegraphics[scale=0.30, trim=0.0cm 0.0cm 0.0cm 0.0cm,clip=true]{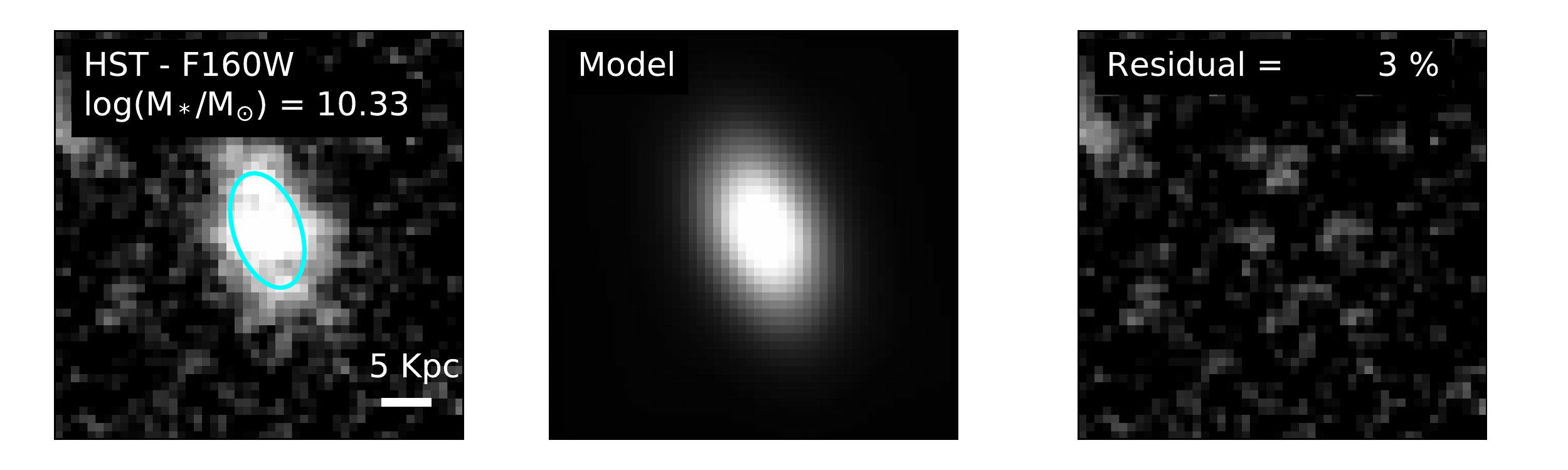}
\includegraphics[scale=0.30, trim=0.0cm 0.0cm 0.0cm 0.0cm,clip=true]{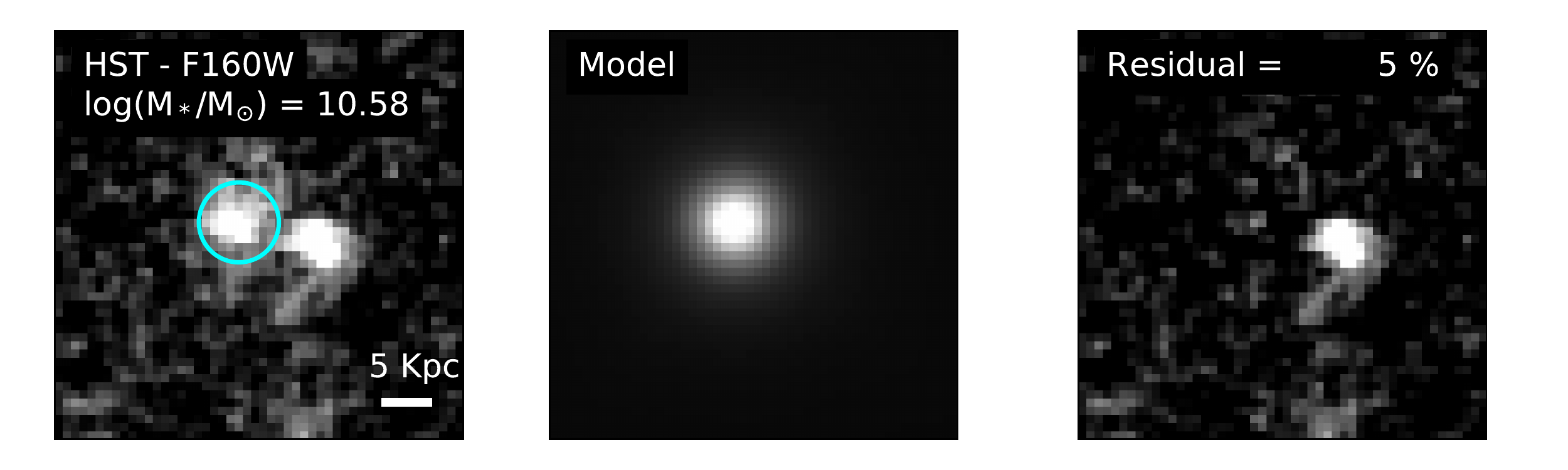}

\caption{Same as Figure \ref{fig:galfit_out_ex} but for all the 10 galaxies with \logmstar\,$>10.2$.  The HST-F160W image and \galfit\ models highlighted with green and red boxes correspond to galaxies with unphysical dynamical masses and possible AGN contaminant, respectively.}
\label{fig:galfit_out_massive}
\end{figure*}

\end{document}